\documentclass[12pt]{iopart}
\usepackage{calc,pstricks,graphicx,iopams,booktabs,,tabularx,setstack}
\usepackage[numbers,square]{natbib}

\newcommand{\fom}{f.o.m.}
\newcommand{\foms}{f.o.m.'s}
\newcommand{\foma}{Triple Detection Rate} 
\newcommand{\fashort}{[3DR]}
\newcommand{\fomb}{Sky Coverage} 
\newcommand{\fbshort}{[SC]}
\newcommand{\fomc}{Directional Precision} 
\newcommand{\fcshort}{[DP]}

\newcommand{\ap}{antenna pattern}
\newcommand{\app}{antenna power pattern}
\newcommand{\aapp}{antenna amplitude pattern}

\newcommand{\D}{D_V}
\newcommand{\DL}{D_{V,L}}
\newcommand{\degree}{${}^o$}

\begin{document}

\title[Networks of gravitational wave detectors]{Networks of gravitational wave detectors and three figures of merit}
\author{Bernard F. Schutz$^{1,2}$}
\address{${}^1$ Albert Einstein Institute, Potsdam, Germany}
\address{${}^2$ School of Physics and Astronomy, University of Cardiff, Wales, UK}
\ead{bernard.schutz@aei.mpg.de}
\date{}

\begin{abstract}
This paper develops a general framework for studying the effectiveness of networks of interferometric gravitational wave detectors and then uses it to show that enlarging the existing LIGO-VIRGO network with one or more planned or proposed detectors in Japan (LCGT), Australia, and India brings major benefits, including much larger detection rate increases than previously thought. I focus on detecting bursts, i.e.\  short-duration signals, with optimal coherent data-analysis methods. I show that the polarization-averaged sensitivity of any network of identical detectors to any class of sources can be characterized by two numbers -- the {\em visibility distance} of the expected source from a single detector and the minimum signal-to-noise ratio (SNR) for a confident detection -- and one angular function, the antenna pattern of the network. I show that there is a universal probability distribution function (pdf) for detected SNR values, which implies that the most likely SNR value of the first detected event will be 1.26 times the search threshold. For binary systems, I also derive the universal pdf for detected values of the orbital inclination, taking into account the Malmquist bias; this implies that the number of gamma-ray bursts associated with detected binary coalescences should be 3.4 times larger than expected from just the beaming fraction of the gamma burst. Using network antenna patterns, I propose three figures of merit that characterize the relative performance of different networks. These measure (a) the expected rate of detection by the network and any sub-networks of three or more separated detectors, taking into account the duty cycle of the interferometers, (b) the isotropy of the network antenna pattern, and (c) the accuracy of the network at localizing the positions of events on the sky.  I compare various likely and possible networks, based on these figures of merit. Adding {\em any} new site to the planned LIGO-VIRGO network can dramatically increase, by factors of 2 to 4, the detected event rate by allowing coherent data analysis to reduce the spurious instrumental coincident background. Moving one of the LIGO detectors to Australia additionally improves direction-finding by a factor of 4 or more. Adding LCGT to the original LIGO-VIRGO network not only improves direction-finding but will further increase the detection rate over the extra-site gain by factors of almost 2, partly by improving the network duty cycle. Including LCGT, LIGO-Australia, and a detector in India gives a network with position error ellipses a factor of 7 smaller in area and boosts the detected event rate a further 2.4 times above the extra-site gain over the original LIGO-VIRGO network. Enlarged advanced networks could look forward to detecting three to four hundred neutron star binary coalescences per year.
\end{abstract}
\pacs{95.55.Ym,95.45.+i}
\submitto{\CQG}

%
%
%
%

\makeatletter
\let\jnl@style=\rm
\def\ref@jnl#1{{\jnl@style#1}}

\def\aj{\ref@jnl{AJ}}                   
\def\actaa{\ref@jnl{Acta Astron.}}      
\def\araa{\ref@jnl{ARA\&A}}             
\def\apj{\ref@jnl{ApJ}}                 
\def\apjl{\ref@jnl{ApJ}}                
\def\apjs{\ref@jnl{ApJS}}               
\def\ao{\ref@jnl{Appl.~Opt.}}           
\def\apss{\ref@jnl{Ap\&SS}}             
\def\aap{\ref@jnl{A\&A}}                
\def\aapr{\ref@jnl{A\&A~Rev.}}          
\def\aaps{\ref@jnl{A\&AS}}              
\def\azh{\ref@jnl{AZh}}                 
\def\baas{\ref@jnl{BAAS}}               
\def\bac{\ref@jnl{Bull. astr. Inst. Czechosl.}}
\def\caa{\ref@jnl{Chinese Astron. Astrophys.}}
\def\cjaa{\ref@jnl{Chinese J. Astron. Astrophys.}}
\def\icarus{\ref@jnl{Icarus}}           
\def\jcap{\ref@jnl{J. Cosmology Astropart. Phys.}}
\def\jrasc{\ref@jnl{JRASC}}             
\def\memras{\ref@jnl{MmRAS}}            
\def\mnras{\ref@jnl{MNRAS}}             
\def\na{\ref@jnl{New A}}                
\def\nar{\ref@jnl{New A Rev.}}          
\def\pra{\ref@jnl{Phys.~Rev.~A}}        
\def\prb{\ref@jnl{Phys.~Rev.~B}}        
\def\prc{\ref@jnl{Phys.~Rev.~C}}        
\def\prd{\ref@jnl{Phys.~Rev.~D}}        
\def\pre{\ref@jnl{Phys.~Rev.~E}}        
\def\prl{\ref@jnl{Phys.~Rev.~Lett.}}    
\def\pasa{\ref@jnl{PASA}}               
\def\pasp{\ref@jnl{PASP}}               
\def\pasj{\ref@jnl{PASJ}}               
\def\rmxaa{\ref@jnl{Rev. Mexicana Astron. Astrofis.}}%
\def\qjras{\ref@jnl{QJRAS}}             
\def\skytel{\ref@jnl{S\&T}}             
\def\solphys{\ref@jnl{Sol.~Phys.}}      
\def\sovast{\ref@jnl{Soviet~Ast.}}      
\def\ssr{\ref@jnl{Space~Sci.~Rev.}}     
\def\zap{\ref@jnl{ZAp}}                 
\def\nat{\ref@jnl{Nature}}              
\def\iaucirc{\ref@jnl{IAU~Circ.}}       
\def\aplett{\ref@jnl{Astrophys.~Lett.}} 
\def\apspr{\ref@jnl{Astrophys.~Space~Phys.~Res.}}
\def\bain{\ref@jnl{Bull.~Astron.~Inst.~Netherlands}} 
\def\fcp{\ref@jnl{Fund.~Cosmic~Phys.}}  
\def\gca{\ref@jnl{Geochim.~Cosmochim.~Acta}}   
\def\grl{\ref@jnl{Geophys.~Res.~Lett.}} 
\def\jcp{\ref@jnl{J.~Chem.~Phys.}}      
\def\jgr{\ref@jnl{J.~Geophys.~Res.}}    
\def\jqsrt{\ref@jnl{J.~Quant.~Spec.~Radiat.~Transf.}}
\def\memsai{\ref@jnl{Mem.~Soc.~Astron.~Italiana}}
\def\nphysa{\ref@jnl{Nucl.~Phys.~A}}   
\def\physrep{\ref@jnl{Phys.~Rep.}}   
\def\physscr{\ref@jnl{Phys.~Scr}}   
\def\planss{\ref@jnl{Planet.~Space~Sci.}}   
\def\procspie{\ref@jnl{Proc.~SPIE}}   

\let\astap=\aap
\let\apjlett=\apjl
\let\apjsupp=\apjs
\let\applopt=\ao
\makeatother

\section{Introduction: detector networks}\label{sec:intro}
\subsection{Current and future networks of interferometers}\label{sec:thenetworks}

The three large gravitational wave detectors of the LIGO project \cite{2009LIGOStatusShort}, located at two sites, and the large instrument  of the VIRGO project \cite{2008VIRGOStatusShortForm}, all of which are expected to reach their Advanced level of sensitivity around 2016, represent the bare minimum required to realize the potential of gravitational wave astronomy when detecting signals of short duration. Using gravitational wave information alone, it is necessary to have at least three separated detectors for locating such sources on the sky, measuring the intrinsic amplitude and polarization of the incoming waves \cite{SCHUTZ1991a}, and determining distances to ``standard-siren'' coalescing compact-object binaries \cite{BBHSirens}. For long-duration (continuous-wave) signals, a single detector can use the phase modulation imprinted by the motion of the Earth to locate sources on the sky. But if the signal is a ``burst'', too short for modulation to be measurable, then positions must be inferred by time-delay triangulation among at least three separated detectors. Some of the most important expected signals will be bursts, such as those from inspiraling and coalescing binaries of neutron stars and/or black holes.

If one of these delicate interferometers temporarily falls out of observing mode or experiences a period of unusually high noise, so that one of the three sites has no functioning detector, or if an incoming gravitational wave arrives from a location on the sky or with a polarization where one of the detectors is significantly less sensitive, then an observation by the remaining detectors will not be able to reconstruct the event completely unless there is other associated information, for example from a gamma-ray burst. Although two-detector observations can have enough significance to identify an event and measure important physical parameters, such as the stellar masses in a binary system, the aim of building detector networks is to extract the greatest possible information from the weak and infrequent signals that we expect to observe with Advanced detectors, and this requires all three sites operated by LIGO and VIRGO.

Fortunately, this network will be enlarged on a short timescale. Funding has begun for the LCGT detector in Japan. There are further proposals for construction in Australia and India. Detectors in Asia or Australia help to cover sky gaps and operational down-times of the basic three and bring an added bonus of improved angular resolution, by increasing the length and number of baselines among detectors of the network. There have been a number of detailed studies of the observing benefits brought by one or another detector \cite{Searle:2002,Arnaud:2003zq,Searle.Networks.2006,2010PhRvD..81h2001W, LIGOsouth,Klimenko2011,0004-637X-725-1-496}. These studies usually simulate network detection by using Monte-Carlo techniques, which provide reliable comparisons of specific configurations but little insight into what would happen with other configurations. It would be helpful, therefore, to have general results applicable to all networks as well as complementary and easily computed ways of quantifying the extra science brought by one or more further detectors. To this end I suggest here three relatively simple figures of merit (\foms) that measure the mean performance of different network configurations. They compare networks' overall event rates (including allowance for realistic duty cycles of detectors), the isotropy of their joint antenna patterns, and the precision with which the networks can measure sky positions of sources. I also derive two general probability distributions for events detected by any network: their observed signal-to-noise ratios, and the observed values of the inclination angle of detected binary systems. The Nissanke {\em et al} \cite{0004-637X-725-1-496} Monte-Carlo study of coalescing-binary detection by various networks is particularly close to the subject of this paper and will provide a useful reference comparison for various analytic results derived below. 

\subsection{Network coherent analysis}\label{sec:netcohere}

The analysis in this paper assumes that a number of detectors observe gravitational waves {\em coherently}, by combining their data in the most sensitive way. The earliest detailed study for gravitational waves of what we now call coherent detection was by G\"ursel and Tinto \cite{GurselTinto1989}. The papers that placed coherent network detection on a sound statistical basis were by Flanagan and Hughes \cite{Flanagan:1997kp} and by Finn \cite{Finn:2000hj}. In this paper I shall concentrate on detecting short-duration signals whose waveform is known in advance, using matched filtering. Coherent detection can also be used to find signals whose waveform is not known \cite{2005PhRvD..72l2002K}.

Coherent detection is not at present the default method of data analysis. All the searches carried out so far by the LSC-VIRGO collaboration have involved coincidence thresholding, which means selecting for further study only stretches of data that appear to contain signals strong enough to pass a pre-determined threshold in two or more detectors, where the signals occur within a maximum time-separation equal to the light-travel time between the detectors (the coincidence ``window''). The experience of current searches has been that most large events in the individual detector data streams are random instrumental artifacts (sometimes called ``glitches''), and the coincidence test eliminates almost all of them because the glitches are not correlated in the data streams of separated detectors. But thresholding is not the optimal signal detection method against Gaussian noise, and in fact it can be very far from optimum, as discussed in \sref{sec:cc} below. Thresholding is used because, although most of the noise background in detectors is Gaussian, glitches make the background far from Gaussian at amplitudes above a few standard deviations. Interferometer-network searches that use thresholding extend methods originally developed for networks of bar antennas \cite{2010PhRvD..82b2003AShort}.

However, networks containing three or more detectors -- our subject in this paper -- have a degree of redundancy that allows them to veto glitches: once the time-delays allow identification of the location of the source, the two polarization waveforms are over-determined by the three or more detector responses. This means that such networks have linear combinations of detector outputs that contain no gravitational wave signal, often called {\em null streams} \cite{GurselTinto1989,WenSchutz2005,2008IJMPD..17.1095W, Klimenko:2008fu}. These can be used to test for and veto glitches, which do not in general cancel out in the null streams. 

Current searches for short-duration signals often follow the thresholding/coincidence step by doing a coherent analysis of the coincident events, in order to use the null-stream vetoes and to extract as much information from them as possible \cite{2010PhRvD..81j2001AShort,PhysRevD.82.102001Short}. In fact, the very first analysis of gravitational wave data from a network of interferometers -- the so-called ``Hundred Hour Run'' -- applied a two-detector null-stream method (after thresholding) to eliminate glitches and show that the strongest observed coincident event had a high probability of occurring by chance \cite{Nicholson1996}, and consequently that no gravitational wave event had been observed.

But the glitch vetoes provided by null streams in principle allow three-detector networks to do fully coherent analysis, without prior thresholding. A number of studies have therefore explored fully coherent detection or compared it with coincidence thresholding \cite{
Arnaud:2003zq,
Klimenko2011,
2008IJMPD..17.1095W,
Klimenko:2008fu,
Pai:2000zt,
2006PhRvD..74h3005M,
2006CQGra..23.4799M,
Chatterji:2006nh,
2007PhRvD..75h7306T,
2007PhRvD..75f2004R,
2008CQGra..25r4025M,
2008CQGra..25r4016H,
2009PhRvD..80l3019M,
2009CQGra..26d5003P,
2010PhRvD..81f2003V}. It is now clear that coherent detection is already able to discriminate real gravitational waves from glitches even in a general three-detector network, and when there are four or more detectors this gets even better. 

Since we will see below that coherent methods are capable of detecting far more events than coincidence methods, it seems reasonable to assume that fully coherent detection will become the norm as the detector network grows. This will not be entirely trivial: one of the principal challenges of introducing coherent data analysis is that it is very demanding of computing, because one has to do a signal search for each resolvable location on the sky. But the payoffs will be worth the effort, especially with the computing power that can be expected to be available by the time the current network is enlarged. The purpose of this paper is, therefore, to characterize the performance of different possible networks when they use coherent detection. 

\subsection{Assumptions and principal results}\label{sec:summary}

The detection sensitivity of a detector network is a function of the sensitivity of the individual detectors and their placement on the earth. An important part of the sensitivity is the network's {\em antenna pattern}, which defines up to a radial scaling the region of space around the earth within which a source should be detected. The overall scale depends on the sensitivity of the individual detectors and the detection threshold that is set for discriminating real signals from noise impersonators. It is conventional in the literature to combine threshold and sensitivity into a radial measure called the horizon distance, the maximum distance a detector or network can detect an event, allowing for an optimum alignment. In this paper I separate threshold from sensitivity by measuring the sensitivity of a detector or network to a given source in terms of a {\em visibility distance}, which is the distance at which the given source would produce a mean response with a signal-to-noise ratio 1, averaged over polarizations. 

From the properties of the antenna pattern I define the three new \foms, for a rather general source population. The \foms\ are meant to be simple to compute and to use. They should give a broad-brush characterization of the effectiveness of networks, but they won't be precise enough to make fine discriminations between similar networks. Although the \foms\ can in principle be computed for any network, I will keep the discussion in this paper simple by making some assumptions.
\begin{itemize}
\item {\bf Detectors.} All the detectors are interferometers with identical sensitivity and identical duty cycles. The detectors' noise streams are not correlated with one another, nor are the times when they drop out of observing mode. The generalization to detectors with different sensitivity is not difficult.
\item {\bf Networks.} The networks are made up of combinations of the Advanced upgrades of the existing LIGO and VIRGO instruments plus planned and possible instruments at the locations in Japan, Australia, and India that are given in \tref{tab:detectors} below. Only networks containing three or more detectors in different locations are considered, because, as noted above, fewer detectors do not return sky position and polarization information from an observation unless there are associated detections in, say, gamma or optical observatories.
\item {\bf Sources.} The gravitational waves all come from an identical population that are randomly and uniformly distributed in (Euclidean) space and in polarization. The waves are short bursts, in that the detectors do not move significantly during the observations, and they are emitted at random times. The waveforms are identical except that they have different overall amplitudes, inversely proportional to the distance to the source; they all have the same polarization evolution (as a function of time) except for a random rotation in the plane of the sky at the start of the burst. Note that, according to this definition, binary systems with different inclinations to the line of sight (different amounts of elliptical polarization) are members of different populations, but binaries with the same inclination but different orientations (rotations in the plane of the sky: the angle $\psi$ in \fref{fig:polConvention}) are members of the same population. We do not consider stochastic signals or long continuous-wave signals from GW pulsars. 
\item {\bf Analysis.} The data are analyzed coherently with a matched filter family capable of matching the incoming signal perfectly. The data analysis finds the ideal match by maximizing the log likelihood. Detector noise is purely Gaussian, at least at the times when events arrive.
\end{itemize}

Given these assumptions, I summarize here the principal results of this paper:
\begin{enumerate}
\item The sensitivity of a network to a population of identical but randomly oriented and randomly located sources depends on the signal waveform, the sensitivity of the detectors (all assumed the same), and the geometry of the network. The signal and sensitivity contribute only to a scaling factor that multiplies the antenna pattern, which depends only on the network geometry \eref{eqn:SNRnetavg}. Therefore the {\em relative} performance of any two networks of similar detectors observing any given source population is independent of the nature of the source. This allows us to compare the advantages and disadvantages of networks without needing to specify much about the signal.
\item The population of detected events has a {\em universal signal-to-noise distribution}, with a probability density function (p.d.f.) proportional to $\rho^{-4}$ above the detection threshold, where $\rho$ is the amplitude signal-to-noise ratio (SNR). The p.d.f.\ \eref{eqn:snrpdf} depends only on the detection threshold $\rho_{\rm min}$ set on $\rho$, not on the geometry or sensitivity of the network.
\item From this p.d.f\ it is possible to deduce that the median amplitude SNR of any detected population will be $2^{1/3} \simeq 1.26$ times the detection threshold $\rho_{\rm min}$. As we wait for the first detection, this is the most likely SNR of the first event, provided that coherent data analysis is used for the search. Similarly, the mean amplitude SNR of the detected population will be 1.5 times the threshold.
\item Binaries with different inclinations have different maximum detection ranges, which biases the expected observed distribution of inclinations. I compute the {\em universal probability distribution for detected inclinations}, independent of the network configuration \eref{eqn:pdfiota}. It peaks around $\pm30^o$. 
\item From this distribution of inclinations one can also deduce another bias, namely that -- provided that mergers involving neutron stars give rise to narrow-beamed gamma-ray bursts -- the number of gamma-ray bursts that will be detected in association with gravitational wave signals will be 3.4 times larger than one would expect if there was no correlation between burst direction and the maximum-power direction of a binary.
\item The first figure of merit (\fom) is called \foma\ \fashort\ (\sref{sec:fomrate}). It measures the rate at which a network can detect events in detectors at three or more separated locations. The rate at which events of a given source population are detected depends of course on the detection volume accessible to the network, but it also depends on the {\em duty cycle} of detectors, which is the fraction of time they spend in observation mode. The first introduction of figures of merit into the discussion of networks seems to have been by Searle, {\em et al} \cite{Searle:2002}, who defined a measure of detection rate that depends effectively only on the detection volume. (See also Searle, {\em et al} \cite{Searle.Networks.2006}.) However, especially at the beginning of the operation of Advanced Detectors, the duty cycle of the detectors will not be 100\%. For the full reconstruction of information about the source, we require at least three separated detectors to observe the event, so a three-detector subnet of a larger network can still return detections. Therefore, \foma\ as defined here is designed to compute how many events can be detected by sub-networks of three or more separated detectors, even when some other detectors in the network may be off the air. 
\item The second \fom\ is called \fomb\ \fbshort\ (\sref{sec:fomisotropy}). It measures the isotropy of the network's antenna pattern. It is defined as the fraction of the $4\pi$ sphere that is covered by the network's antenna pattern at a range that is $1/\sqrt2$ of the maximum. For a given number of detectors of a standard sensitivity, there is a trade-off between isotropy and overall detection volume: if the antenna patterns of individual detectors reinforce each other, then the volume they include will be larger than if they fill in each other's directional ``holes''. But isotropy might be a desirable thing in itself. For example, if the source population is anisotropic (perhaps biased toward the Galactic plane) then an isotropic network might do better than one with a larger range. Or if the sources are expected to be associated with objects that can be detected also by a non-gravitational signal, but only if they are relatively nearby compared to the maximum range of the network (e.g.\ supernovae seen with neutrinos), then an isotropic network could do better. 
\item The third \fom\ is called \fomc\ \fcshort\ (\sref{sec:fomaccuracy}). It measures how well the network localizes events on the sky, its directional accuracy. Generally speaking, longer baselines improve direction-finding. \fomc\ uses the measure of solid-angle error introduced by Wen and Chen \cite{2010PhRvD..81h2001W}. It is proportional to an average over the antenna pattern, not of the size of the error box, but of its inverse. This prevents the measure being distorted by small regions where direction-finding is poor; instead it is weighted more by the regions of the sky where direction-finding is particularly good.
\item Enlarging the basic LIGO-VIRGO network with detectors in Japan and/or Australia also provides a less obvious but perhaps even more important benefit: it makes coherent data analysis more robust and allows the detection of events that would not pass the coincidence threshold tests used in the current LIGO-VIRGO data analysis (\sref{sec:cc}), where fully coherent analysis is difficult because of the geometry of the network. This could lead to an improvement of as much as a factor of 4 in the recovery of signals within a given detection volume, depending on the effectiveness with which coherent methods can be introduced into the data analysis of the basic LIGO-VIRGO network.
\end{enumerate}

By comparing these measures for various possible networks, some simple lessons emerge. First, if one takes as a baseline the performance of the original network of Advanced detectors -- LIGO Hanford with two full-size interferometers, LIGO Livingston, and VIRGO -- using coherent detection, then there is a big win in event rate from putting another large detector anywhere in Asia or Australia. This comes partly from adding more detection volume and partly from providing greater coverage when individual detectors randomly drop out of observing mode. A Japanese detector (LCGT) makes the antenna pattern more isotropic; an extra Australian detector (AIGO) makes its reach go deeper. If instead of building an extra detector in Australia, one of the LIGO Hanford instruments is placed in Australia (LIGO Australia), the improvement in detection rate is not quite as dramatic. The big change then is a significant improvement in direction-finding. If we take the network that includes LIGO Australia and LCGT, again there is a very big improvement in the event rate, and of course it becomes more isotropic as well. This is pretty much a ``dream configuration'' in terms of present opportunities. If a project gains traction in India and a large INDIGO detector is built, then this produces even further gains that the \foms\ quantify. On top of these improvements due to detector numbers and geometry, the robustness of coherent detection in an enlarged network will lead to further striking gains in event rate over the current coincidence style of analysis.

It is important to remark that these \foms\ should be regarded as rules of thumb, not as exact measures of the performance of any network. But treated with a small amount of caution, the measures show how big the science gains can be from adding further Advanced detectors to the existing three sites.

\section{Network antenna patterns and the amplitude distribution of detected events}

\subsection{Antenna pattern and detection volume of a single interferometer}\label{sec:app}

\begin{figure}
\centering
\includegraphics[width=3in]{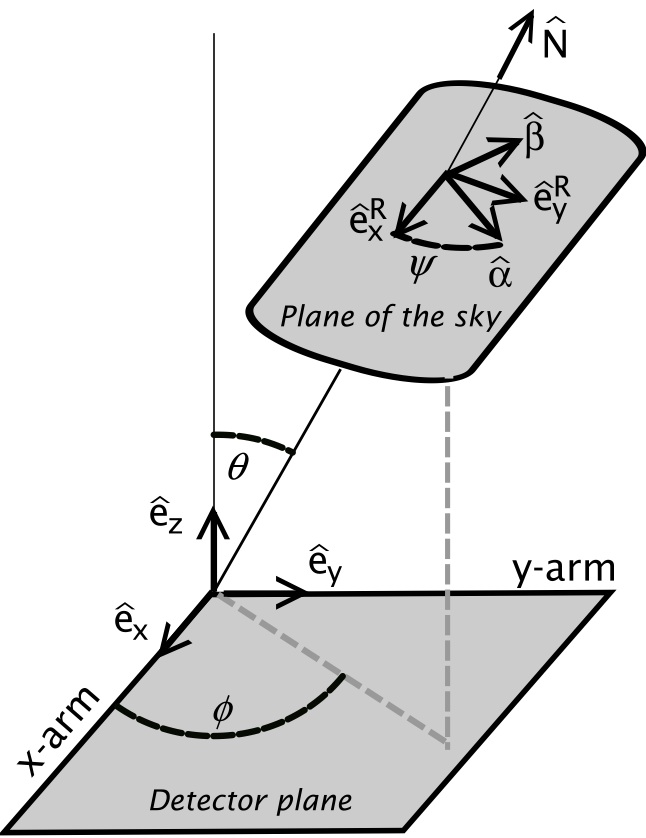}
\caption{The relative orientation of the sky and 
detector frames. From \cite{SathyaSchutz:LivingReviews}.}
\label{fig:polConvention}
\end{figure}

The \foms\ are based on the antenna patterns of the detectors, which describe their relative sensitivity in different directions. Each detector is linearly polarized and has a quadrupolar antenna pattern. In the notation of Sathyaprakash and Schutz \cite{SathyaSchutz:LivingReviews}, we consider a detector in the $x-y$ plane with arms along the axes, and a gravitational wave coming from a direction given by the usual spherical coordinates $\theta$ and $\phi$ relative to the detector's axes, whose two polarization components $h_+$ and $h_\times$ are referred to axes in the plane of the sky that are rotated by an angle $\psi$ relative to the detector axes (see \fref{fig:polConvention}, which is taken from figure~3 of Sathyaprakash and Schutz \cite{SathyaSchutz:LivingReviews}). Then the strain $\delta L/L$ of the interferometer is 
\begin{equation}\label{eqn:interferometerantenna}
\frac{\delta L(t)}{L} = F_+(\theta,\,\phi,\,\psi) h_+(t) + 
F_\times(\theta,\,\phi,\,\psi) h_\times(t), 
\end{equation}
where the $F_+$ and $F_\times$ are the \ap\ functions for the 
two polarizations. Using the geometry in \fref{fig:polConvention},
one can show that 
\begin{eqnarray}\label{eqn:interferometerantennapatterns}
F_+ & = & \frac{1}{2}\left (1+\cos^2\theta \right )\cos 2\phi \cos 2\psi - 
\cos\theta\sin 2\phi \sin 2\psi, \nonumber \\
F_\times & = & \frac{1}{2}\left (1+\cos^2\theta \right )\cos 2\phi \sin 2\psi + 
\cos\theta\sin 2\phi \cos 2\psi.
\end{eqnarray}

These are the \ap\ response functions of the interferometer to the two polarizations of the wave as defined in the sky plane \cite{Thorne1987}. Note that the maximum value of both $F_+$ and $F_\times$ is 1.

Sometimes the angle $\eta$ between the arms of a detector is not exactly $\pi/2$, for reasons of local geography or by design. For that reason it is helpful to orient the detector in the $x$-$y$ coordinate plane by aligning the {\em bisector} of the angle between the arms with the bisector of the angle between the axes \cite{schutz.tinto}. One also has to multiply the functions $F_+$ and $F_\times$ in \eref{eqn:interferometerantenna} by $\sin\eta$. When we discuss networks we will define the orientation of the detector to be the geographical direction of the arm bisector.

The expected power signal-to-noise ratio (SNR) of the signal in the detector's data stream is, if it can be discovered by ideal matched filtering, 
\begin{equation}\label{eqn:SNR}
\rho^2 = 4\int_0^\infty \frac{|\tilde{\delta L}(f)/L|^2}{S_h(f)}{\rm d}f,
\end{equation}
where $S_h(f)$ is the one-sided spectral noise density normalized to the gravitational wave amplitude, and the time-series strain $\delta L(t)/L$ in \eref{eqn:interferometerantenna} has been Fourier-transformed into $\tilde{\delta L}(f)/L$, which then depends on the Fourier transforms $\tilde{h}_+(f)$ and $\tilde{h}_\times(f)$ of the incoming waves. I will assume from now on that we are detecting a short burst of gravitational waves, so that the detector does not change its orientation during the observation. A discussion of network detection of long-duration signals, such as those from gravitational wave pulsars, may be found in Cutler and Schutz \cite{2005PhRvD..72f3006C,2007PhRvD..75b3004P}.

\begin{figure}[htbp]
\begin{center}
\includegraphics[width=3in]{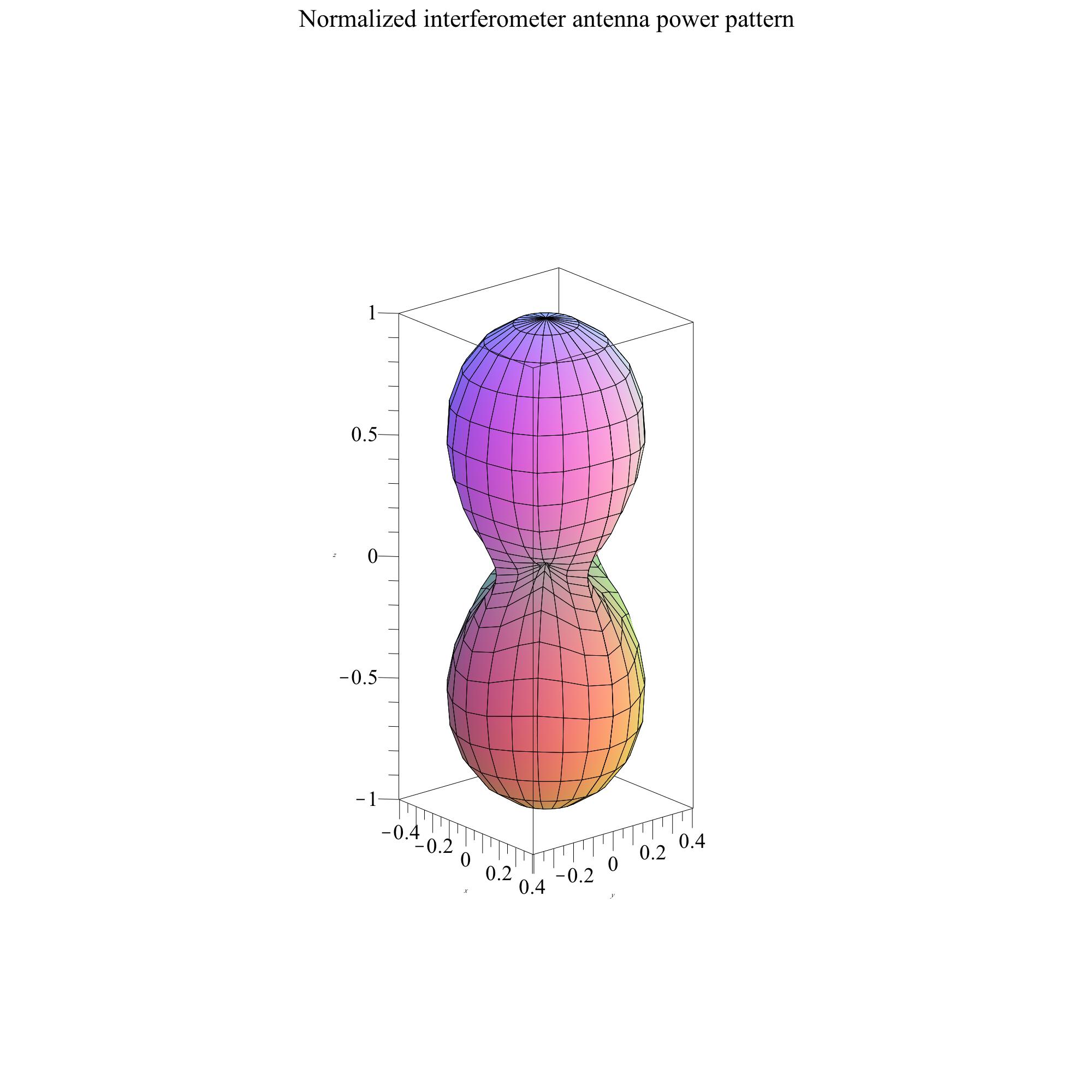}
\includegraphics[width=3in]{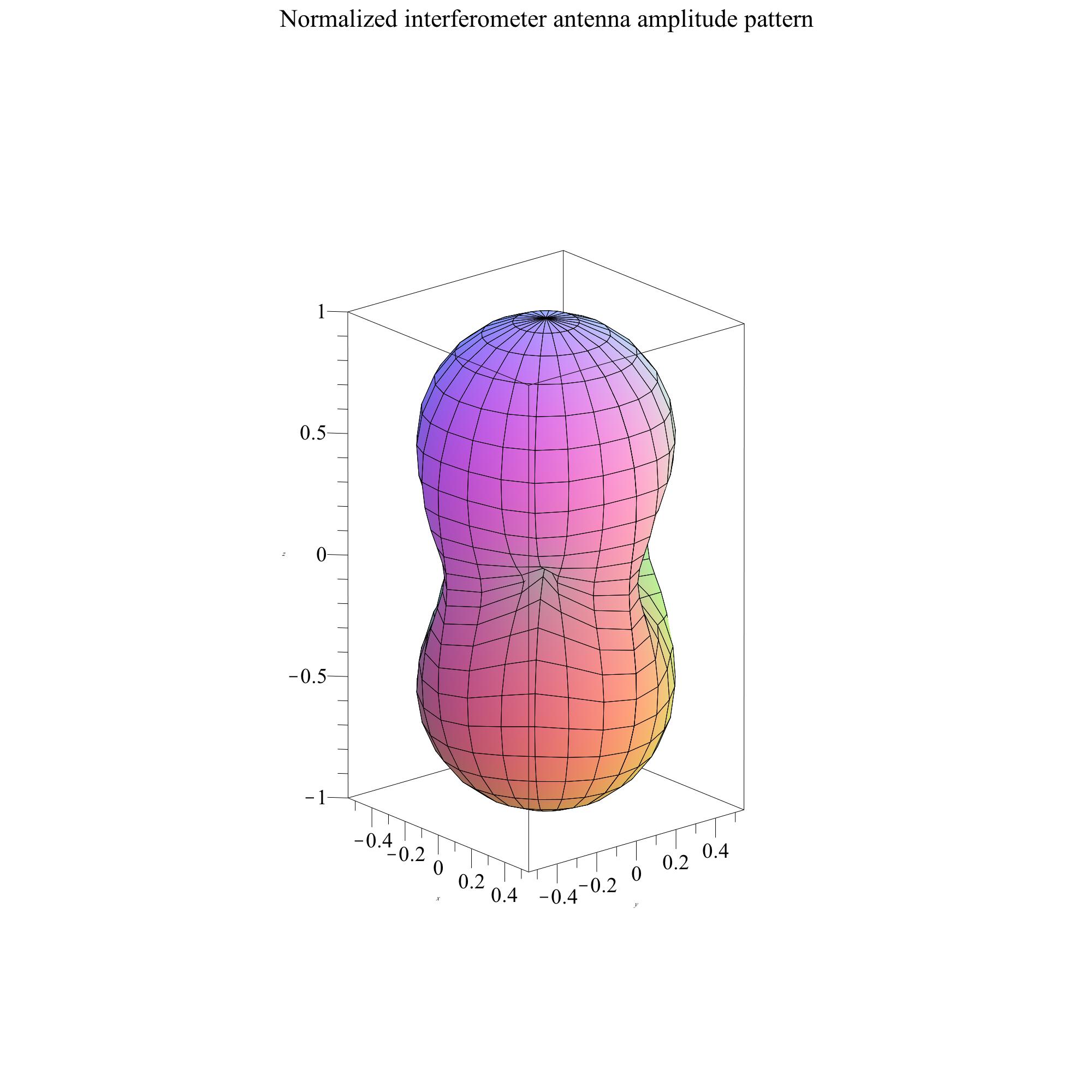}
\caption{The antenna power pattern (left panel) and its square-root (amplitude pattern: right panel) of a single interferometer oriented with axes in the $x$-$y$ plane, averaged over polarizations of the incoming wave. The amplitude pattern represents the shape of the detection volume of the instrument, or its maximum detection reach in different directions.}
\label{fig:peanut}
\end{center}
\end{figure}

We now apply the assumption that the wave has a randomly oriented polarization. Consider a source which emits wave components $H_+(f)$ and $H_\times(f)$, referred to its own frame, defined perhaps by some  preferred axis or plane in the source. Suppose that at the {\em start} of the observation this source frame is different from the detector frame as projected onto the sky by a rotation angle $\alpha$. During the observation the polarization will rotate in some way determined by $H_+(f)$ and $H_\times(f)$. This is of no interest to us here. The important point is that the ensemble of sources at the same position in space contains systems with all possible {\em initial} angles $\alpha$. When we average the power SNR in \eref{eqn:SNR} over the ensemble, we will simply be changing in a uniformly random way the projection of the source's intrinsic $+$ and $\times$ components onto the detector's. The result is that the mean power SNR over the ensemble (denoted by $\left<\;\right>$) depends only on the sum of the squares of the sensitivity functions of the detector to both polarizations:
\begin{equation}\label{eqn:SNRavg}
\left<\rho^2\right> = 2 \left[F_+(\theta,\phi,\psi)^2 + F_\times(\theta,\phi,\psi)^2\right]\int_0^\infty\frac{|H(f)|^2}{S_h(f)}{\rm d}f,
\end{equation}
where $|H(f)|^2 = |H_+|^2 + |H_\times|^2$. We call the function
\begin{eqnarray}
P(\theta,\phi) &=& F_+(\theta,\phi,\psi)^2 + F_\times(\theta,\phi,\psi)^2\nonumber\\
&=&\frac{1}{4}(1+\cos^2\theta)^2\cos^22\phi + \cos^2\theta\sin^22\phi\label{eqn:app}
\end{eqnarray}
the {\em antenna power pattern} of a single interferometer. Note that, from \eref{eqn:interferometerantennapatterns}, the antenna power pattern is independent of the angle $\psi$ that is the reference angle for the wave's polarization, as one would expect after our ensemble polarization average. It is plotted in the detector coordinate frame in \fref{fig:peanut}. This is often referred to as the ``peanut diagram''.

If, for a single detector, there is a detection threshold $\rho_{\rm min}$ on the amplitude SNR, then a signal from a direction $(\theta,\,\phi)$ can be expected to be detected if 
\begin{equation}\label{eqn:threshold}
2P(\theta,\phi)\int_0^\infty\frac{|H(f)|^2}{S_h(f)}{\rm d}f \ge \rho_{\rm min}^2.
\end{equation}
For the purposes of our discussion, we suppose that the gravitational wave source has a standard intrinsic amplitude, so that its received amplitude $H(f)$ is inversely proportional to the distance $r$ to the source. We also suppose that these sources are randomly distributed in space. Let us normalize the amplitude by defining (arbitrarily) a standard reference distance $r_s$ at which our source would have amplitude $H_s(f)$, so that a source at a distance $r$ has amplitude
\begin{equation}\label{eqn:refdist}
H(f) = \frac{r_s}{r}H_s(f).
\end{equation}
This is much the way astronomers distinguish between absolute and apparent magnitudes, by defining the absolute magnitude to equal the apparent magnitude of the source if it were at a fixed fiducial distance (10 pc). 

Explicitly separating $r$ out in $H(f)$ will be helpful for the volume integrals below. For example, we can now rewrite \eref{eqn:SNRavg} as
\begin{equation}\label{eqn:SNRavgRewrite}
\left<\rho^2\right> = \frac{2}{r^2}P(\theta,\phi)\int_0^\infty\frac{|r_sH_s(f)|^2}{S_h(f)}{\rm d}f.
\end{equation}
Note that the product $rH=r_sH_s$ is independent of the distance to the source. We use this to define the {\em visibility distance} of the source $\D$:
\begin{equation}\label{eqn:visibility}
{\D}^2 = 2\int_0^\infty\frac{|r_sH_s(f)|^2}{S_h(f)}{\rm d}f.
\end{equation}
This is the distance at which the source would have SNR = 1 in a single detector at its most sensitive location in the sky, namely directly overhead at $\theta=0$ or $\pi$, after averaging over the sky polarization angle $\psi$. All the details of filtering and the detector noise curve are hidden in the single parameter $\D$. This leads to a simple way of writing \eref{eqn:SNRavgRewrite}
\begin{equation}\label{eqn:SNRsimple}
\left<\rho^2\right> = P(\theta,\phi)\frac{{\D}^2}{r^2}.
\end{equation}

Similarly, I will define the {\em mean horizon distance} $R_0$ for a single detector observing this source to be the distance at which the source is on average just at the detection threshold $\rho_{\rm min}$ when it is overhead, so that $R_0=\D/\rho_{\rm min}$. Then the {\em reach} $R$ of the single detector in any direction ${\theta,\phi}$ is 
\begin{equation}\label{eqn:reach}
R(\theta,\phi) = R_0 [P(\theta,\phi)]^{1/2} = \frac{\D}{\rho_{\rm min}}[P(\theta,\phi)]^{1/2}.
\end{equation}
We call the square-root of the \app\ the {\em \aapp}. The volume bounded by the reach $R(\theta,\phi)$ is called the {\em detection volume}. Its size is determined by the \aapp, scaled by the mean horizon distance $R_0$. The mean horizon distance is smaller than what is conventionally called the horizon distance, which is the distance at which an optimally {\em polarized} source exactly overhead can just be detected at threshold.

Note that we are making an approximation here when we define a detection volume by polarization averaging. Sources at the edge of the volume have only a 50\% chance of being detected, while those that are well inside are detected with higher probability. Moreover, a number of sources outside this volume will be detected if they have a favorable polarization.  Our approximation is to replace the real detection probability distribution in space with a fixed volume that has a hard edge: everything inside is detected, everything outside is missed. We use this approximation only to study the gross properties of detection, such as numbers detected, typical position accuracies, and so on, and only to compare different networks. The comparisons are likely to be better than the accuracy of the approximation for any single network, since the errors will systematically affect all networks the same way. The test of how accurate this approximation is for any specific network is whether it matches up with Monte-Carlo studies of the real detection problem for that network.

\subsection{Antenna pattern of a network of detectors}\label{sec:appnet}

We now need to generalize these concepts to networks of more than one detector. I will assume here that the detectors' noise streams are uncorrelated. This is a good assumption for all networks except those that include two detectors at the Hanford LIGO site. Even there, experience has shown that the correlations can be reduced to a very low level with careful experimental design. A full treatment of the theory of detection in networks that have detectors with correlated noise may be found in Finn \cite{Finn:2000hj}.

When computing the joint antenna pattern of the entire network, the antenna patterns of the individual detectors must of course be transformed to a common celestial coordinate system. We take this to be the Earth-based spherical coordinates, and from now on we denote them by $(\theta,\phi)$. In addition, there must be a common definition of the incoming wave polarization. I use here the formulation given in \cite{JKS1998}, whose expressions were developed for the problem of long-term observations, where the detector changes orientation with time. For the present paper we merely need to set $t=0$ in their formulation, and we shall use conventional spherical sky coordinates rather than declination and right-ascension.

As shown by Finn \cite{Finn:2000hj}, the network power SNR is just the sum of the power SNRs of the individual detectors
\begin{equation}\label{eqn:SNRnet}
\rho^2_N = \sum_{k=1}^{N_D}\rho^2_k,
\end{equation}
where $N_D$ is the number of detectors and where we define the individual power SNRs as
\begin{equation}\label{eqn:SNRcomponent}
\rho^2_k = 2\int_0^\infty\frac{|H_k(f)|^2}{S_{h}(f)}{\rm d}f,
\end{equation}
where $H_k(f)$ is the waveform projected onto the $k$-th detector. Averaging as before over the random polarization angle, we have
\begin{equation}\label{eqn:SNRnetavg}
\left<\rho^2_N\right> = 2\sum_k (F_{+,k}^2 + F_{\times, k}^2)\int_0^\infty\frac{|H(f)|^2}{S_{h}(f)}{\rm d}f,
\end{equation}
where $F_{+,k}$ and $F_{\times, k}$ are the antenna patterns of the individual detectors. Note that the integral in this equation does not depend on $k$ and is therefore taken outside the sum. The sum is then the function
\begin{equation}\label{eqn:appsum}
P_N(\theta,\phi) = \sum_k (F_{+,k}^2 + F_{\times, k}^2),
\end{equation}
which is called the {\em network antenna power pattern}.\footnote{If the detectors were not identical, then one could modify the network antenna pattern simply by including a single weighting factor consisting of the ratio of $\rho^2$ for each detector to a standard detector $\rho^2$ for the particular signal waveform being considered. The network antenna pattern would then be waveform-dependent.} This is our analytic approximation to the detection sensitivity found in \cite{Searle.Networks.2006} from Monte-Carlo studies of randomly oriented binary systems. In terms of $P_N$ the network SNR takes the simple and useful form
\begin{equation}\label{eqn:SNRnetscale}
\left<\rho^2_N\right> = P_N(\theta,\phi)\frac{\DL^2}{r^2},
\end{equation}
where $\DL$ is the visibility distance of a single detector, labelled here as the Livingston detector. Remember, all detectors are assumed identical so all have the same visibility distance. This assumption is easily dropped if necessary, but it makes the discussion in the present paper simpler.

It is worth remarking that the polarization-averaged network antenna pattern does not depend on the local orientation of each detector, since it is the sum of the individual detector power patterns \eref{eqn:appsum}, and for each detector the sum of the squares of the antenna pattern components is invariant under rotations of the detector in its plane. It might seem counterintuitive that two co-located detectors with orthogonal orientation make the same average contribution to the signal power received by the network as they would if they were perfectly aligned. When aligned they work well together but miss many events that one of them would catch when not aligned. When searching for a stochastic gravitational wave signal, of course, alignment is crucial. Moreover, even for bursts, the ability to determine polarization and sky position of a signal will be affected by the relative alignment of the detectors. I will return to this point later.

The resulting expression for the antenna pattern of an arbitrarily located and oriented interferometer in our notation is as follows. The source position is given by the spherical coordinates $(\theta,\phi)$ on the sky, and the frame for the wave polarization angle $\psi$ is defined to be aligned with this spherical-coordinate grid. The detector is at  latitude $\beta$ and longitude $\lambda$. It is an interferometer oriented such that the bisector of its arms points in the direction $\chi$,  measured counter-clockwise from East. Its arms have an opening angle of $\eta$. The celestial coordinates $(\theta,\phi)$ are aligned with latitude and longitude, so that the equators of both systems coincide and the celestial point $(\theta=\pi/2,\,\phi=0)$ is in the zenith direction above the geographic location $(\beta = 0,\, \lambda = 0)$. The antenna pattern functions are
\begin{eqnarray}
F_+ &=& \sin\eta[a\cos(2\psi)+b\sin(2\psi)],\label{eqn:fplusgen}\\
F_\times &=& \sin\eta[b\cos(2\psi)-a\sin(2\psi)]\label{eqn:fcrossgen},
\end{eqnarray}
where the functions $a$ and $b$ are given by
\begin{eqnarray}  
\fl a = \frac{1}{16}\sin(2\chi)[3-\cos(2\beta)][3-\cos(2\theta)]\cos[2(\phi+\lambda)]+\nonumber\\ \frac{1}{4}\cos(2\chi)\sin(\beta)[3-\cos(2\theta)]\sin[2(\phi+\lambda)]+\nonumber\\
\frac{1}{4}\sin(2\chi)\sin(2\beta)\sin(2\theta)\cos(\phi+\lambda)+\nonumber\\ \frac{1}{2}\cos(2\chi)\cos(\beta)\sin(2\theta)\sin(\phi+\lambda)+\frac{3}{4}\sin(2\chi)\cos^2(\beta)\sin^2(\theta),\label{eqn:a}\\
\fl b = \cos(2\chi)\sin(\beta)\cos(\theta)\cos[2(\phi+\lambda)]-\frac{1}{4}\sin(2\chi)[3-\cos(2\beta)]\cos(\theta)\sin[2(\phi+\lambda)]+\nonumber\\ \cos(2\chi)\cos(\beta)\sin(\theta)\cos(\phi+\lambda)-\frac{1}{2}\sin(2\chi)\sin(2\beta)\sin(\theta)\sin(\phi+\lambda).\label{eqn:b}
\end{eqnarray}

\subsection{Detection volume of a network of detectors}\label{sec:detvol}

The {\em detection volume} $V_N$ of the network is defined as the region enclosed by its reach in any direction, which as before is 
\begin{equation}\label{eqn:reachnetwork}
R_N(\theta,\phi) = R_0 [P_N(\theta,\phi)]^{1/2} = \frac{\DL}{\rho_{N,{\rm min}}}P_N^{1/2},
\end{equation}
where $R_0$ is defined as before to be the mean horizon distance (maximum reach) of a single detector for this source at the chosen {\em network} detection threshold SNR $\rho_{N,{\rm min}}$, and where (as before) $\DL$ is the single-detector visibility distance (maximum range at SNR = 1). {\em I will assume that when we compare networks, all of them have the same detection threshold.} 

We can compute the detection volume explicitly: 
\begin{eqnarray}
V_N&=& \int {\rm d}\Omega\int_0^{R_N(\theta,\phi)}r^2dr=\frac{1}{3} \int {\rm d}\Omega R_N^3(\theta,\phi)\nonumber\\ &=& \frac{1}{3}R_0^3 \int {\rm d}\Omega [P_N(\theta,\phi)]^{3/2}.
\label{eqn:detectionvolume}
\end{eqnarray}

\begin{figure}[htbp]
\begin{center}
\includegraphics[width=3in]{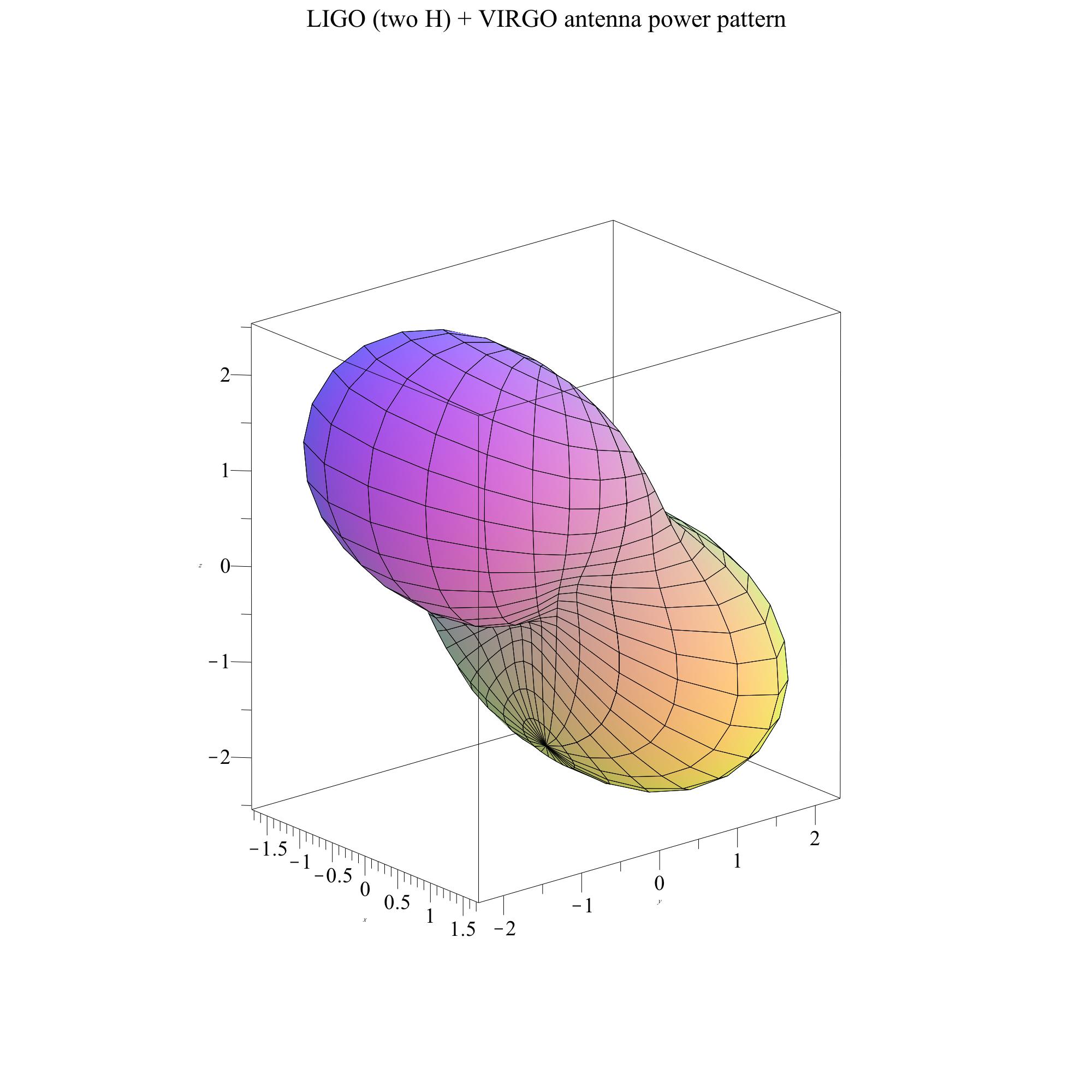}
\includegraphics[width=3in]{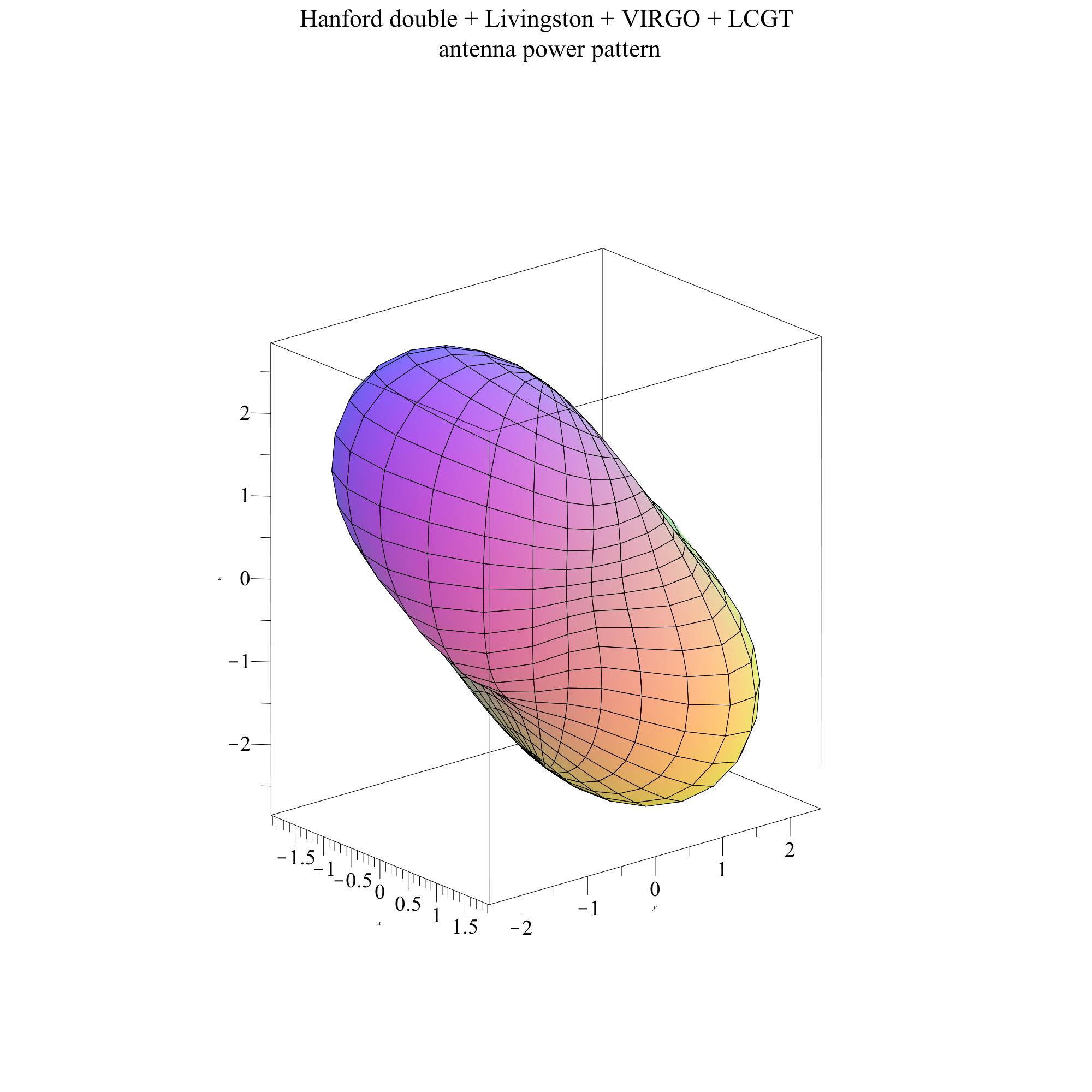}
\includegraphics[width=2.5in]{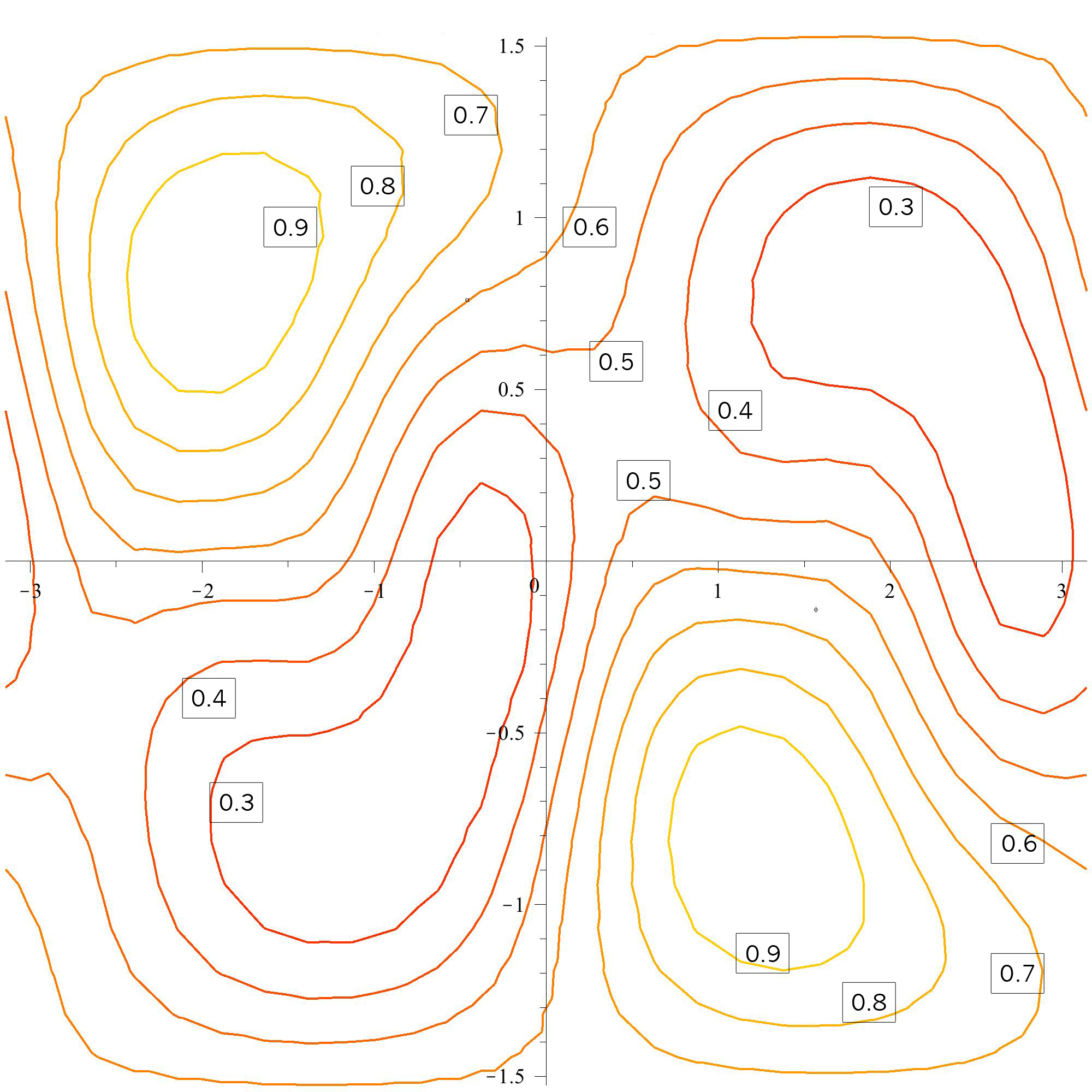}
\includegraphics[width=2.5in]{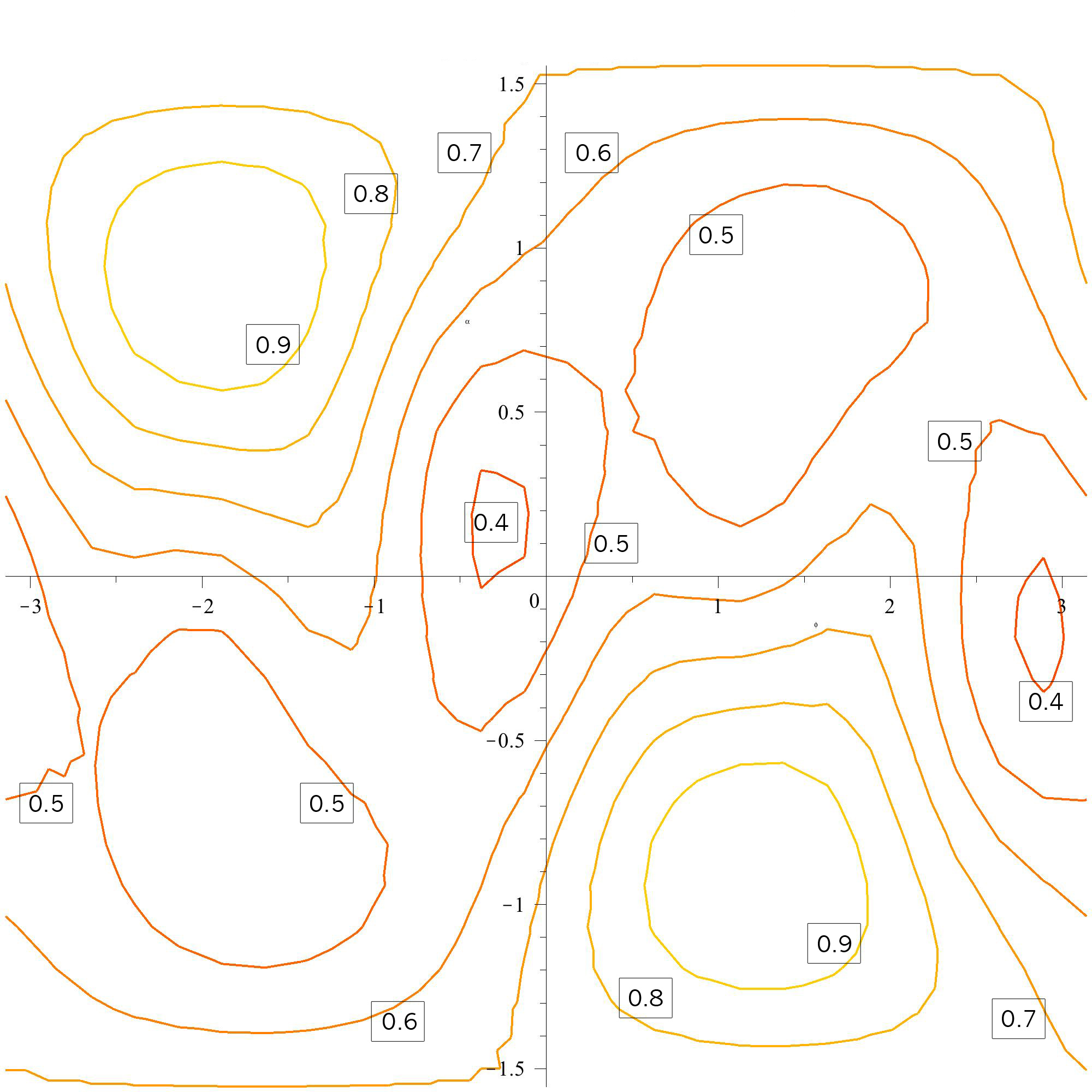}
\caption{The antenna power patterns of the LIGO and VIRGO detector network with two detectors at Hanford (HHLV: left panel) and of the network after including the Japanese detector LCGT (HHJLV: right panel). All detectors are assumed to be identical.  As in Figure \ref{fig:peanut}, the sensitivity is averaged over polarizations of the incoming wave. {\em Top row:} The coordinate system is oriented with $z$ aligned with geographic North and the $x$-axis at geographic longitude 0\degree. In all such plots from now on, the viewer is located at longitude 40\degree W and 20\degree N, above the mid-Atlantic. Note that all antenna patterns are reflection symmetric through the center of the earth, so that the hidden side is a mirror image of the side shown in the diagram. {\em Bottom row:} The same data plotted as contour plots. Contours are labeled with values relative to the maximum. For HHLV on the left, the maximum is 3.03 (square of mean horizon distance from \tref{tab:networks}).  For HHJLV on the right, the maximum is 3.31.}
\label{fig:power.HHLV.HHLVJ}
\end{center}
\end{figure}

\Tref{tab:detectors} gives the important parameters of the detector locations that will be considered in this paper, including the one-letter abbreviation by which the detectors will be denoted in naming the various networks. 

As an illustration, in \fref{fig:power.HHLV.HHLVJ} the network antenna power patterns are plotted for two networks: the planned Advanced network of two LIGO detectors at Hanford, one at Livingston, and VIRGO; and the same network plus the LCGT detector in Japan. Notice that the hole in the southwest direction has been filled in by the Japanese detector.

\subsection{Universal distribution of detected amplitudes}\label{sec:universal}

Because the angular sensitivity of the detectors is totally decoupled from the dependence of SNR on the distance of the source, which resides in $H(f)$ in \eref{eqn:SNRnetavg}, we can work out the expected distribution of SNR for detected events analytically for any detector network and source population. To do this we make explicit in \eref{eqn:detectionvolume} the fact that $R_N$ is inversely proportional to the detection threshold $\rho_{N,{\rm min}}$, by using \eref{eqn:reachnetwork}:
\begin{equation}\label{eqn:detvolthresh}
V = \frac{\DL^3}{3\rho_{N,{\rm min}}^{3}} \int {\rm d}\Omega [P_N(\theta,\phi)]^{3/2}.
\end{equation}
The number of detections with SNR larger than any given $\rho_N$ is proportional to the detection volume with $\rho_{N,{\rm min}}$ set equal to this $\rho_N$. This scales as $\rho_N^{-3}$. This is a cumulative distribution: the number of detections with SNR larger than $\rho_N$ scales as $\rho_N^{-3}$. It is straightforward from this to show that the {\em universal probability density function for the distribution of detected} SNR {\em values} is 
\begin{eqnarray}\label{eqn:snrpdf}
p(\rho_N){\rm d}\rho_N &={} 3\rho^3_{N,{\rm min}}\rho_N^{-4}{\rm d}\rho_N, &\qquad \rho > \rho_{N,{\rm min}}\\
&={} 0, &\qquad \rho_N < \rho_{N,{\rm min}}. \nonumber
\end{eqnarray}

From this simple universal distribution one can deduce any of the moments one wishes. For example, the {\em mean expected amplitude SNR} is $1.5\rho_{N,{\rm min}}$. The {\em mean expected power SNR}${}^2$ is $3\rho_{N,{\rm min}}^2$. 

The median of this distribution is of particular interest and can also be deduced from a simple argument: it is the value of the threshold for which the detection volume is one-half of the full volume. Since the volume scales as the inverse cube of the threshold, the {\em median amplitude SNR} value will be $2^{1/3}\rho_{N,{\rm min}}$. The {\em median power SNR}${}^2$ is $2^{2/3}\rho_{N,{\rm min}}^2$. The importance of the median is that it is the most likely SNR value of the first signal that will be detected. It has often been remarked that the rapid increase of volume with distance means that the first source is likely to be near the detection limit. Here we quantify that statement: {\em the most likely amplitude SNR of the first detection is $2^{1/3}\simeq1.26$ times the threshold of the search.} The median source is weaker than either the amplitude mean or the power mean. That is because the universal distribution has a peak at the lowest values (at threshold) and has a long tail of strong but rare events. 

Of course, this argument has been made in the context of our antenna pattern detection criterion, which is an approximation.  However, I believe one can expect that the distribution should be close to the distribution of real observations, provided the detection criterion depends on coherent addition of signals against mainly Gaussian noise.

\subsection{Detection volumes for binary systems}\label{sec:binarydetvol}

As remarked in the definition of sources in \sref{sec:intro}, binary systems with different inclinations belong to different source populations as far as our detection volumes are concerned, because the strength of their emitted radiation depends on inclination, and their own radiation patterns are anisotropic; in fact, if we were to average the power pattern shown in the peanut diagram  (\fref{fig:peanut}) over circles around its long axis we would get a plot of the radiation power pattern of a binary system. But binaries with different inclinations are all members of the same physical family, just seen from different and random directions. Therefore it is interesting here to consider binary detection as a function of inclination angle $\iota$.

\begin{figure}[htbp]
\begin{center}
\includegraphics[width=3in]{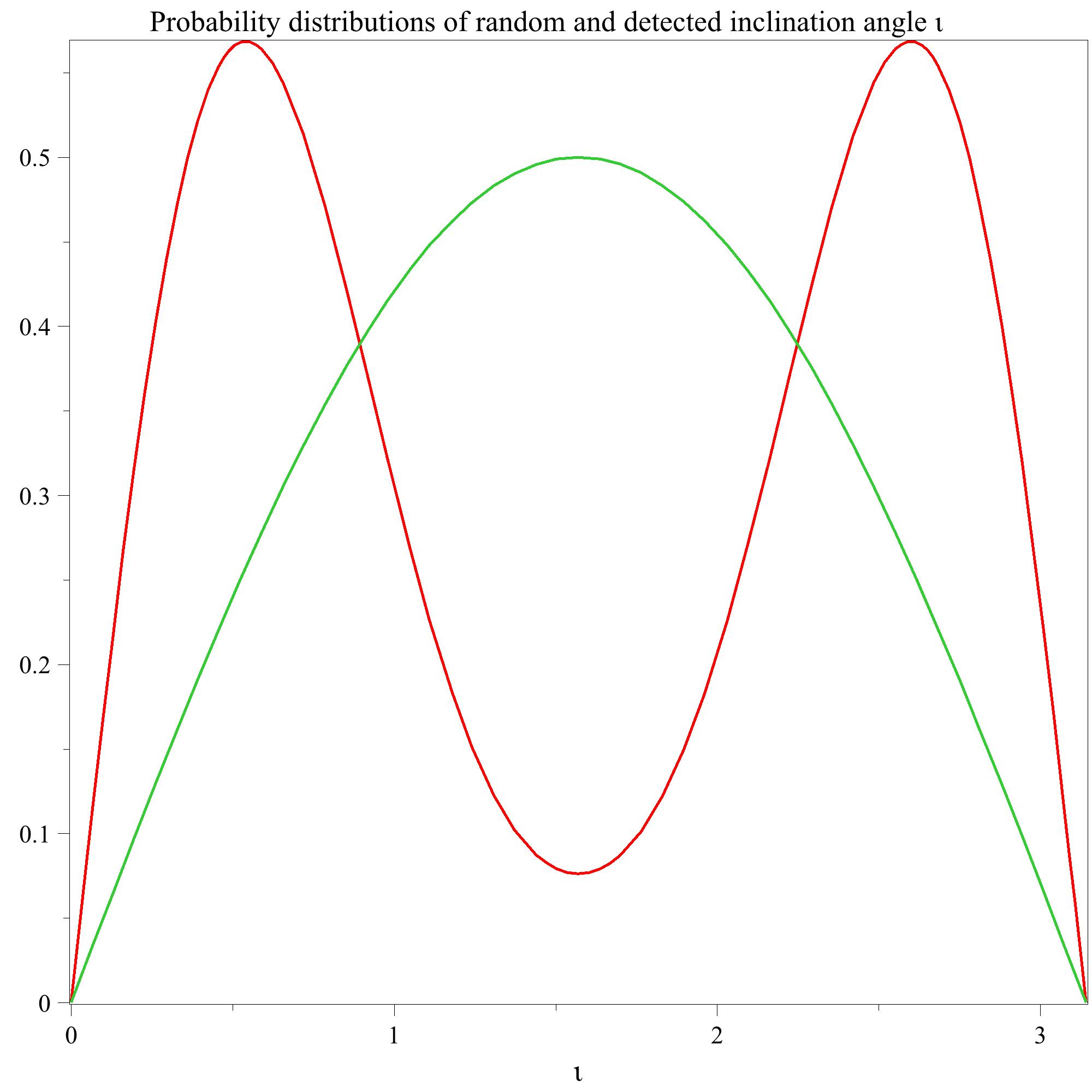}
\caption{The probability distributions of inclination angle $\iota$ (in radians) for randomly oriented binaries (the single-peaked curve, which is just $\sin\iota$) and for detected binaries (the double-peaked curve, from \eref{eqn:pdfiota}). The selection bias (essentially the Malmquist bias) toward low inclinations due to the anisotropic radiation pattern of a binary is clear.}
\label{fig:iotabias}
\end{center}
\end{figure}

The maximum power is radiated along the rotation axis of the binary, defined as $\iota=0$, and the minimum power in its orbital plane, $\iota=\pi/2$. For a general inclination angle it is easy to show from, e.g., Sathyaprakash and Schutz \cite{SathyaSchutz:LivingReviews} that the radiated power depends on inclination in the following way:
\begin{equation}\label{eqn:radiationinclination}
P_{{\rm rad}}(\iota)= F_{\rm rad}(\iota)P_{{\rm rad}}(\iota=0),
\end{equation}
with 
\begin{equation}\label{eqn:radiationpattern}
F_{\rm rad}(\iota)=\frac{1}{8}(1+6\cos^2\iota + \cos^4\iota).
\end{equation}
We call this function the {\em binary radiation pattern}. As remarked above, this is the $\phi$-average of the interferometer's antenna pattern \eref{eqn:app}. The detection volume will depend on $\iota$ as 
\begin{equation}\label{eqn:detectioninclination}
V_{N}(\iota)=\left[F_{\rm rad}(\iota)\right]^{3/2}V_{N}(\iota=0),
\end{equation}
This predicts the relative numbers of sources that will be detected, i.e.\  it quantifies the bias toward small inclination angles created by the stronger radiation pattern in those directions. We can derive the probability distribution function of detected values of $\iota$ by normalizing $F_{\rm rad}^{3/2}$ over the intrinsic distribution of angles, which has the probability distribution function $\sin\iota$. The normalizing integral is 
\[\int_0^\pi \left[F_{\rm rad}(\iota)\right]^{3/2}\sin\iota\,{\rm d}\iota = 0.58092.\]
The probability distribution of detected values of $\iota$ is therefore 
\begin{equation}\label{eqn:pdfiota}
p_{\rm det}(\iota)=0.076076(1+6\cos^2\iota + \cos^4\iota)^{3/2}\sin\iota.
\end{equation}
This is plotted in \fref{fig:iotabias}. Note that this, also, is a universal distribution, in that it applies to any network doing coherent analysis. As with the distribution of detected values of SNR, this result is exact only within the approximation we are making that the polarization-averaged antenna pattern defines a detection volume with a sharp boundary. This pdf is completely consistent with the Monte-Carlo result of Nissanke {\em et al} \cite{0004-637X-725-1-496} (their figure 3), when allowance is made for the difference between using $\iota$ as the independent variable (here) and $\cos\iota$ (\cite{0004-637X-725-1-496}).

The mean value of $V_{N}(\iota)$ in \eref{eqn:detectioninclination} is $0.29046V_{N}(\iota=0)$. This means that the expected number of binaries detected, allowing for random inclination and polarization angles, is about 29\% of the number that would be expected if all the systems were face-on. 

\Fref{fig:iotabias} also has implications for coincidences between gravitational wave detections and gamma-ray bursts. If we accept the popular model in which a coalescence of two neutron stars or a neutron star and a black hole is accompanied by a gamma-ray burst that is emitted in a narrow cone around the binary's rotation axis, then events where the cone points toward us are also stronger gravitational wave emitters, and so we will see relatively more of them \cite{1993ApJ...417L..17K}. The slope of the distribution of detected binaries in \fref{fig:iotabias} at $\iota=0$ is about 1.72, compared with 0.5 for the true distribution, a ratio of 3.44. Therefore, a coincidence between a gravitational wave event and a gamma burst with a narrow cone (so that only the linear behavior of the curves in the figure is relevant) is about 3.4 times more likely than one would expect by just naively computing the solid angle of the jet. For example, if jets have a solid angle of $4\pi/100$, then only one out of every one hundred coalescences would point its jet toward us. But we could expect that one in every 29 {\em detected} coalescences would be accompanied by a gamma-ray burst.

\section{Figures of merit}\label{sec:foms}
\subsection{\foma: Relative effectiveness of a network}\label{sec:fomrate}

The first of the figures of merit measures the relative effectiveness of a network at detecting the short bursts of gravitational waves that we assume in our signal model, using enough detectors to extract the full information available in the gravitational wave signal. Since all detectors are assumed identical and the source waveform is the same in each case, only the network detection volume and the duty cycle need to be used to provide a realistic measure of the relative rates at which events will be detected by different networks.

The relative detection volumes of various networks calibrate the volume of space accessible to the network (often given in current LSC-Virgo papers in units of MWEG: Milky Way Equivalent Galaxies). But adding extra detectors to a network does more than increase its detection volume. It also ensures that there is less time when there are fewer than three detectors in operational mode. Current interferometers need exquisitely tuned control systems to keep the interferometry locked on a fringe. During the recent S5 science run \cite{2009PhRvD..80j2001A}, the two big LIGO detectors achieved a duty cycle of about 80\%. When the detectors start up at the advanced level of sensitivity, around 2016, the duty cycle may well be similar. In principle there is no reason that the duty cycle could not ultimately be pushed well above 90\%, but this will require time and effort. (The smaller GEO600 detector achieved a 95\% duty cycle during S5 and VIRGO operated at close to that efficiency during its several-month participation at the end of S5.) If one requires an observation to be performed by all instruments in a three-detector network with a duty cycle of 80\% then they will be observing simultaneously only $0.8^3\simeq51\%$ of the time. If one adds a fourth detector, the amount of time {\em at least three detectors} will be in observing mode dramatically increases to $(0.8)^4+4(0.2)(0.8)^3\simeq82\%$. Adding a fifth raises this to $(0.8)^5+5(0.2)(0.8)^4+10(0.2)^2(0.8)^3 \simeq 94\%$, a further significant increase. We can expect that these numbers will be realistic during the first few years of the operation of Advanced detectors, until the experimental teams can focus their efforts on improving duty cycle instead of raw sensitivity.

{\em The \foma\ figure of merit for a given network sums the detection volumes of all sub-networks containing detectors in three or more locations, each weighted by the probability that the given sub-network will be the only one observing at a given time.} We do not include the amount of time that only two detectors are in operation because these cannot fully reconstruct the event in the absence of other information. Specifically, then, consider a network of 4 separated detectors, called A, B, C, and D, all of which are in observing mode for a fraction $f$ of the data-taking time, and whose down-times are not correlated with one another. We define the \foma\ of this network to be the effective available volume, with the scaling factor $(\DL/\rho_{N,{\rm min}})^3$ removed:
\begin{eqnarray} \label{eqn:fomvol}
\fl \fashort_{\rm ABCD} &=& \left(\frac{\DL}{\rho_{N,{\rm min}}}\right)^{-3}\left[ f^4V_{\rm ABCD} + (1-f)f^3(V_{\rm ABC}+V_{\rm BCD}+V_{\rm ACD}+V_{\rm ABD})\right],\nonumber\\
&=& \frac{f^4}{3}\int {\rm d}\Omega [P_{ABCD}(\theta,\phi)]^{3/2} + \frac{(1-f)f^3}{3} \int {\rm d}\Omega\left\{ [P_{ABC}(\theta,\phi)]^{3/2} +\right.\nonumber\\ && \left. [P_{BCD}(\theta,\phi)]^{3/2} +  [P_{ACD}(\theta,\phi)]^{3/2} +  [P_{ABD}(\theta,\phi)]^{3/2} \right\}.
\end{eqnarray}
\foma\ is thus a measure of the {\em effective three-site detection volume} averaged over a long observing run. The number of events detected by three or more separated detectors in a network during a given observing period will be proportional to the network's value of \foma. The generalization of \eref{eqn:fomvol} to networks with other numbers of detectors is obvious.

The definition of \foma\ specifies detectors at different sites because a network of 3 detectors involving two at Hanford cannot resolve sky positions, and hence cannot infer polarizations, distances, and other parameters. Therefore, in computing $\fashort_{\rm HHLV}$, the original four-detector Advanced network, I do not use \eref{eqn:fomvol}. Instead of all four three-detector subnetworks, I include only two, both having the antenna power pattern HLV, but involving different Hanford detectors. With this assumption and an 80\% duty cycle, we get $\fashort_{\rm HHLV}=4.86$. This serves as a reference value for other networks, since it is the basic coverage available from the presently funded Advanced detectors with a realistic duty cycle for the initial operation. 

By contrast, if one of the Hanford detectors is placed in Australia, we get the network AHLV, which has $\fashort_{\rm AHLV}= 6.06$ with a duty cycle of 80\%. The rate of events whose locations can be measured goes up by 25\% simply by separating the two Hanford detectors, because doing this creates two more useful three-detector sub-networks. On the other hand, with a 95\% duty cycle, the difference is not so pronounced: $\fashort_{\rm HHLV}=7.81$ while $\fashort_{\rm AHLV}= 8.28$. In this case, most detections occur with all four detectors working, for which in both configurations there is always a subset of three at separate locations. We return to compare other interesting specific networks in \sref{sec:applications} below.

To convert \fashort\ back to an effective detection volume in space, multiply by $(\DL/\rho_{N,{\rm min}})^3$, where $\DL$ is the visibility distance of the source for the Livingston detector (the distance at which an optimally located source has unit SNR), and $\rho_{N,{\rm min}}$ is the network detection threshold SNR. To convert this effective volume into an expected detection rate one multiplies by the volume rate of events of this population.

\subsection{Isotropy}\label{sec:fomisotropy}
If the \ap s of detectors in a network are well-aligned, they increase the detection volume nonlinearly, since the detection volume of a small solid angle in any direction depends on the $3/2$ power of the total \app. Where the \ap s do not overlap significantly, they make the network more isotropic. Increasing the detection volume is obviously an important gain, but there may also be merit in a network that is more isotropic. Isotropic \ap s are better for coincidence observations with other all-sky survey instruments, particularly those that are significantly flux-limited with a range shorter than that of the gravitational wave detectors, as for example neutrino detectors searching for gravitational collapse events \cite{2008CQGra..25k4051AShortForm,0264-9381-27-8-084019}. In such a coincidence observation the events will be relatively nearby, so the isotropy of the antenna pattern is more important than its total volume.  This illustrates the key point that the importance attached to different values of the \foms\ depends on one's priorities in building a new detector, a point also made in \cite{Searle.Networks.2006}.

{\em We define the \fom\ \fomb\ to be the fraction of the sky over which the network's \app\ is greater than half of its maximum value.} By cutting the sky at this value we are accepting all directions where the reach of the network is at least $1/\sqrt{2}\simeq71\%$ of its mean horizon distance $R_N$. The concept of sky coverage was discussed for single detectors in Sathyaprakash and Schutz \cite{SathyaSchutz:LivingReviews}, but the sky cut was done there at 50\% of the mean horizon distance. The place where the cut is made is clearly arbitrary, but since detection is based on computing SNR${}^2$, I use the 50\% power level in defining this \fom.

Networks differ greatly in their isotropy. For a single interferometer, \fbshort\ is just 34\%. Aligning antenna patterns keeps them anisotropic, so networks including the LIGO detectors and an Australian detector tend to have low values of \fbshort, while adding in VIRGO or LCGT increases isotropy. Again, this is illustrated for specific interesting networks in \sref{sec:applications}. The AHJLV network, with detectors in Australia and Japan, reaches 85\%, and adding a detector in India pushes the sky coverage over 90\%.

\begin{table}[htbp]
   \centering
   \begin{tabular}{lcccc} 
   	\toprule
      Detector & Label & Longitude & Latitude & Orientation\\
	\midrule
      LIGO Livingston, LA & L & 90\degree\ 46' 27.3" W & 30\degree\ 33' 46.4" N & 208.0\degree (WSW) \\
      LIGO Hanford, WA & H & 119\degree\ 24' 27.6" W & 46\degree\ 27' 18.5" N & 279.0\degree (NW) \\
      VIRGO, Italy & V & 10\degree\ 30' 16" E & 43\degree\ 37' 53" N &  333.5\degree(NNW) \\
      LCGT, Japan & J & 137\degree\ 10'  48" E & 36\degree\ 15' 00" N & 20.0\degree (WNW)  \\
      AIGO, Australia & A & 115\degree\ 42' 51" E  & 31\degree\ 21' 29" S & 45.0\degree (NE) \\
      INDIGO, India & I & 74\degree\ 02' 59" E & 19\degree\ 05' 47" N & 270.0\degree (W) \\
      \bottomrule
   \end{tabular}
   \caption{Name, abbreviation, geographic location, and orientation of the various detector positions considered in this paper. The abbreviations will be used to label functions and diagrams. When there are two instruments at Hanford we will use HH. The orientation is the geographic compass angle, measured clockwise from North, of the line bisecting the arms of the detector. (This decouples the orientation from opening angle for detectors that may not have perpendicular arms.) For the averages performed in this paper, however, the orientation will not matter. The data for the LIGO and VIRGO detectors are for the actual detectors. The data for LCGT are for the planned orientation. The data for AIGO are from the Australian group (private communication) and place the detector at Gin-gin. The data for INDIGO are essentially arbitrary; they correspond to the location of GMRT and an arbitrary orientation. Opening angles $\eta$ are not listed because all detectors are assumed to have $\eta = \pi/2$.}
   \label{tab:detectors}
\end{table}

\subsection{Accuracy}\label{sec:fomaccuracy}

The biggest benefit of adding one or more detectors in Asia or Australia is that they add longer baselines to the existing three detectors, and it is the baseline that determines the accuracy with which the source can be located on the sky. Source resolution is achieved by time-delay triangulation, so that for fixed errors in measuring the time-of-arrival of a signal at different detectors, longer baselines provide better relative accuracy and smaller sky-position errors. Position accuracy in turn affects the determination of other parameters: if the position is wrong then the inferred intrinsic amplitude of the signal and its polarization will be wrong. This issue has been studied for specific networks, particularly those containing a detector in Australia, which offers the longest baselines \cite{2010PhRvD..81h2001W,2009NJPh...11l3006F, 2010FairhurstLocalization}. These studies sometimes provide detailed sky maps of error ellipses under various assumptions, and they show that for any network the angular resolution varies considerably over the sky. The purpose here is instead to develop a single measure that captures the general difference in resolution when one compares two different networks. The \fom\ called \fomc\ attempts to provide a simple sky-averaged measure of the relative accuracy with which a given network can determine positions. 

The problem of determining how accurately a network can measure positions has a long history. Triangulation should produce angular position errors proportional to the time-of-arrival measurement error divided by the baseline between two detectors, measured in light-travel time \cite{SCHUTZ1991a}. But since three detectors need to be involved in order to narrow down the position to a single location on the sky (or at most two locations), the geometry of the detector array is key. The first quantitative conjecture on the solid-angle uncertainty for a network of three gravitational wave detectors appeared in G\"ursel and Tinto \cite{GurselTinto1989}, who refer to a private communication by K S Thorne. The geometric characteristic they use is the area $A_\perp$ of the triangle of the detectors projected perpendicular to the direction to the source. The solid angle error $\delta\Omega$ for a source in a particular direction is, according to G\"ursel and Tinto,
\begin{equation}\label{eqn:deltaomegaold}
\delta\Omega = 2\frac{(c\delta t_{12})(c\delta t_{13})}{A_\perp},
\end{equation}
where $\delta t_{12}$ and $\delta t_{13}$ are the rms timing errors on two of the arms of the triangle. This improves when the SNR improves because the timing errors decrease. No proof of this expression seems to have appeared in the literature until the recent work of Wen, Fan, and Chen \cite{2008JPhCS.122a2038W,2010PhRvD..81h2001W}, who give a much more general exact result that reduces to this when the network consists of three identical detectors. I will base \fomc\ on a simplification of the Wen-Fan-Chen expressions, which in their full form allow the exact computation of position errors for networks of any number of non-identical detectors.

Wen and Chen \cite{2010PhRvD..81h2001W} show that the solid angle uncertainty is given by
\begin{eqnarray}\label{eqn:wenchen41}
(\delta\Omega)^{-2} = \frac{\sum_{j,k,\ell,m} \xi_j\xi_k\xi_\ell \xi_m |(\mathbf{r}_{kj} \times \mathbf{r}_{m\ell })\cdot {\mathbf{n}}|^2}{\left[4\sqrt{2}\pi c^2 \sum_j\xi_j\right]^2},
\label{Omega}
\end{eqnarray}
where the sum is over detectors in the network, $\mathbf{n}$ is the direction to the source, and $\mathbf{r}_{kj}$ is the vector from detector $k$ to detector $j$. (It follows that in the sum, $k$ and $j$ are distinct, as are $m$ and $\ell$.) The symbol $\xi_j$ provides the timing accuracy, and for our case, where we assume we can do perfect matched filtering, it is:
\begin{equation}\label{eqn:defxi}
\xi_j = \left<\omega^2\right>_j\rho^2_j = (\delta t_{arr, j})^{-2},
\end{equation}
where $\rho^2_j$ is the squared SNR in detector $j$, where $\left<\omega^2\right>_j$ is the mean squared frequency in the signal, averaged over the signal waveform in the detector weighted inversely by the detector noise, and where $\delta t_{arr, j}$ is the r.m.s.\ time-of-arrival measurement error in detector $j$ when there are no covariances with other measurement errors \cite{SCHUTZ1991a}. Notice that \eref{eqn:wenchen41} depends on the projected areas of all the various triangles formed by the inter-detector vectors. If there are only three detectors, there is only one triangle, and this expression essentially reduces to \eref{eqn:deltaomegaold}.

The measure \eref{eqn:wenchen41} is the inverse of one element of the error covariance matrix, and is therefore an estimate of the inverse of the area of the 1-$\sigma$ error ellipse. It is also related to the Fisher information matrix element for solid angle. My definition of \fomc\ in \eref{eqn:fomcdef} below inherits this: it is to be regarded as an indicator of the 1-$\sigma$ errors in area. This is an important point to bear in mind when comparing with other authors, who often quote 90th percentile or 2-$\sigma$ errors.

If we assume that all detectors are identical, then all the $\left<\omega^2\right>_j$'s are the same and all the $\xi_j$'s are proportional to the squares of their respective detector antenna pattern, multiplied by factors that are common to all detectors. Our first simplification will be to ignore the polarization-dependence of the antenna patterns for the sources and take 
\[\xi_j =  \left<\omega^2\right> P_j D_L^2/r^2.\]
This is not strictly equivalent to taking a polarization average of the solid angle uncertainty, but when using the expression to compare different networks on average this should be a small correction. The sum $\sum_j\xi_j$ is then proportional to the network power pattern $P_N$. 

The next simplification is that I will replace each individual detector power pattern $P_j$ by the average of the network power pattern, $P_N/N_D$. Again this is in the spirit of finding a simple measure associated with the network as a whole. It is equivalent to saying that the network power SNR is equally shared by all detectors.

For the final step we have to decide what it is that we integrate to get a measure of accuracy. Is it appropriate to find a measure of $|\delta\Omega|$, $|\delta\Omega|^2$, $|\delta\Omega|^{-1}$, $|\delta\Omega|^{-2}$, \ldots? Any of these might be useful for comparing different networks. I shall opt for something proportional to (with the previously mentioned simplifications) an average value of $|\delta\Omega|^{-1}$, mainly for reasons of ease of computation. This measure is more sensitive to locations where $|\delta\Omega|$ is small, that is, where the network gives particularly good directional information. An average of $|\delta\Omega|$ itself would be dominated by the regions where directions are poor. Given the relationship between \eref{eqn:wenchen41} and the Fischer information, the measure used here can also be thought of as an indicator (no more than that) of the directional information contained in the measurement: the larger the value of \fomc\, the more directional information we get.  

It follows from these assumptions that 
\begin{equation}\label{eqn:avginvarea}
\left<\left|(\delta\Omega)^{-2}\right|^{1/2}\right> \simeq \frac{R_\oplus^2\left<\omega^2\right>}{4\pi c^2}\rho_{N,{\rm min}}^2\fcshort,
\end{equation}
where I {\em define}, for any network of $N_D$ detectors, the \fomc\ of the network to be
\begin{equation}\label{eqn:fomcdef}
\fcshort = N_D^{-2}(V_N)^{-1}\int {\rm d}\Omega P_N^{3/2} \left[\sum_{k>j,m>\ell } |(\tilde{\mathbf{r}}_{kj} \times \tilde{\mathbf{r}}_{m\ell })\cdot {\mathbf{n}}|^2\right]^{1/2}.
\end{equation}
Here $V_N$ the network's total detection volume, normalized in such a way that a single interferometer has maximum range 1 (\eref{eqn:detectionvolume} with $R_0=1$), $R_\oplus$ is the Earth's radius, and $\tilde{\mathbf{r}}_{kj} = \mathbf{r}_{kj}/R_\oplus$ is the vector connecting the locations of detectors $j$ and $k$ on the unit sphere (i.e.\ in latitude and longitude).  

Larger values of \fcshort\ indicate better direction accuracy. The scale factor in \eref{eqn:avginvarea} evaluates straightforwardly to give
\begin{equation}\label{eqn:fomctoaccuracy}
\left<\left|(\delta\Omega)^{-2}\right|^{1/2}\right>  \simeq 14 \rho^2_{N,{\rm min}}\left(\frac{\left<\omega^2\right>}{(2\pi \times 100 \;\rm Hz)^2}\right)\fcshort\quad\mbox{sr}^{-1}.
 \end{equation}
 
Note that, in the sum over detectors in \eref{eqn:fomcdef}, the sum is restricted to pairs where $k$ exceeds $j$ and $m$ exceeds $\ell$. This is justified because, as noted above, these indices cannot be equal and because including values where $k<j$ would simply count the same detector pair twice. The coefficient in front of the sum has been increased by a factor of $\sqrt{2}$ to compensate. Terms for which $k$ equals $m$ and $j$ equals $\ell$  also vanish because they involve the cross product of a vector with itself. The sum shown therefore has $(N_D+1)N_D(N_D-1)(N_D-2)/4$ nonvanishing terms. This number of terms, inside the square root, is roughly compensated by the factor of $N_D^{-2}$ outside the integral, which arose from our simplification in which we replaced each individual detector power pattern $P_j$ by the average of the network power pattern $P_N/N_D$. The fact that $N_D$ roughly cancels out means that \fcshort\ depends more on the size of the detector triangles than on the number of detectors in the network: extending the baselines in a network has more effect on angular accuracy than does adding more detectors with similar baselines to the existing ones.

It should be noted that \fcshort\ measures the average position accuracy of detected signals, not the accuracy on a given signal with a fiducial amplitude. If network A is more sensitive than network B, so that A has a bigger detection volume, then its position accuracy will be averaged over a population that includes more distant and weaker sources than those of B. If we only asked how network A would perform on the detection volume of network B, its mean direction accuracy would be better than one might guess just by comparing $\fcshort_A$ with $\fcshort_B$. So when using \fcshort\ to compare the performance of different networks, it is somewhat easier to interpret when it is used to compare networks with the same number of detectors but different geometries.

When combined with the values of \fcshort\ we compute in \tref{tab:networks}, this is not inconsistent with the plots of error ellipses in the literature \cite{2010PhRvD..81h2001W,LIGOsouth,2010FairhurstLocalization,Klimenko2011}. The dependence on threshold $\rho_{N,{\rm min}}$ is interesting: the higher the threshold, the stronger the ensemble of detected SNRs, so the larger the value of \fcshort, and the better the direction-finding.

\begin{table}[htbp]
   \centering
   \begin{tabularx}{\linewidth}{lXXXXXXX} 
   	\toprule 
      Network & Mean Horizon Distance & Detection Volume & Volume Filling Factor & \foma\ \mbox{(at 80\%)} & \foma\ \mbox{(at 95\%)} & \fomb & \fomc\\
	\midrule
L               & 1.00 & 	1.23 & 29\% & -      & -       &  33.6\% & - \\
HLV           & 1.43 & 5.76 & 47\% & 2.95 &  4.94 & 71.8\% & 0.68 \\
HHLV        & 1.74 & 8.98 & 41\% & 4.86 &  7.81 & 47.3\%  & 0.66 \\
AHLV        & 1.69 & 8.93 & 44\% & 6.06 &   8.28 & 53.5\% & 3.01 \\
HHJLV      & 1.82 & 12.1 & 48\% & 8.37 & 11.25 & 73.5\% &  2.57 \\
HHILV       & 1.81 & 12.3 & 50\% & 8.49 & 11.42 & 71.8\% &  2.18 \\
AHJLV       & 1.76 & 12.1 & 53\% & 8.71 & 11.25 & 85.0\% &  4.24 \\
HHIJLV      & 1.85 & 15.8 & 60\% & 11.43 & 14.72 & 91.4\% & 3.24 \\
AHIJLV      & 1.85 & 15.8 & 60\% & 11.50 & 14.69 & 94.5\% &  4.88 \\      \bottomrule
   \end{tabularx}
   \caption{Comparison of various networks. A: AIGO or LIGO Australia; H: LIGO Hanford single detector; HH: LIGO Hanford two detectors; I: INDIGO; J: LCGT; L: LIGO Livingston; V: VIRGO. Mean Horizon Distance is the maximum detection distance, scaled to the mean horizon distance (maximum range) of a single detector observing at the same threshold. Detection Volume is the volume inside the antenna pattern, on the same scale. Volume Filling Factor is the ratio between the Detection Volume in column 3 and the volume of a sphere with radius equal to the Maximum Range in column 2. The remaining columns are the figures of merit.  \foma\ measures the overall detection rate and is given for two different values of the duty cycle: 80\% to represent a likely figure at the start of operations, and 95\% to represent a reasonable long-term operation goal.  The values of \foma\ are smaller than the Detection Volume by factors representing the loss of 3-site observing time to duty cycle downtime. \fomb\ measures how isotropic the network antenna pattern is. \fomc\ reflects angular accuracy: the typical solid angle uncertainty is inversely proportional to \fomc, so that larger values denote more accurate networks. The first row of the table is for a single detector, to facilitate comparisons.}
   \label{tab:networks}
\end{table}

\section{Lessons}\label{sec:applications}

\subsection{Discussion of specific networks}\label{sec:nets}

At the present time the only network of Advanced detectors that is fully approved and funded consists of two LIGO detectors at Hanford and one at Livingston, plus VIRGO in Italy: HHLV. Working together, these four detectors have a detection volume of 8.98, more than 7 times that of a single detector at the same network threshold. But when the duty cycle is 80\% the effective volume \fashort\ falls to 4.86. The network covers 47\% of the sky at half-power. Its value of 0.66 for \fcshort\ is the starting point for comparisons of network accuracy.

In addition to these detectors, funding has started for the LCGT detector in Japan, so it is reasonable to expect that the Advanced network will include detectors at Hanford and Livingston in the USA, in Italy, and in Japan. If the current proposal to move one of the Hanford detectors to Australia becomes reality, then we should have the network AHJLV. If not, then we are likely to have HHJLV. In addition, if a proposal to build a detector in India succeeds, then in the long run we could have AHIJLV or HHIJLV. 

To understand the capabilities of these networks it is useful to compare them with the basic HHLV and with LIGO's own variant AHLV. These comparisons will show clearly the considerable benefits brought by the ongoing investment in Japan and the proposed investment in India.

First we ask, how does AHLV compare to HHLV? The range and volume of AHLV are very similar to those of HHLV. Its effective detection rate, \fashort, however, is 25\% larger: 6.06 (compared to 4.86) at a duty cycle of 80\%, simply because there are more three-site sub-networks in this array. AHLV is slightly more isotropic than HHLV, with \fbshort\ equal to 53.5\%. This reflects the fact that the position of the Australian detector at Gingin is very close to being antipodal to the LIGO detectors. So far these network characteristics are not very different from HHLV. But the real improvement is in direction finding. The value of \fcshort\ for AHLV is 3.01, compared with 0.66 for HHLV. This suggests that the typical error ellipses will be reduced in area by more than 4 if the detector is moved to Australia. These numbers are consistent with the results of the much more extensive comparison of these two networks in an unpublished internal technical report of the LIGO Scientific Collaboration \cite{LIGOsouth} and in a recent study of coherent detection involving LIGO Australia \cite{Klimenko2011}, and they give a very strong scientific reason for placing the LIGO instrument in Australia, independently of other detector developments.

\begin{figure}[htbp]
\begin{center}
\includegraphics[width=2.4in]{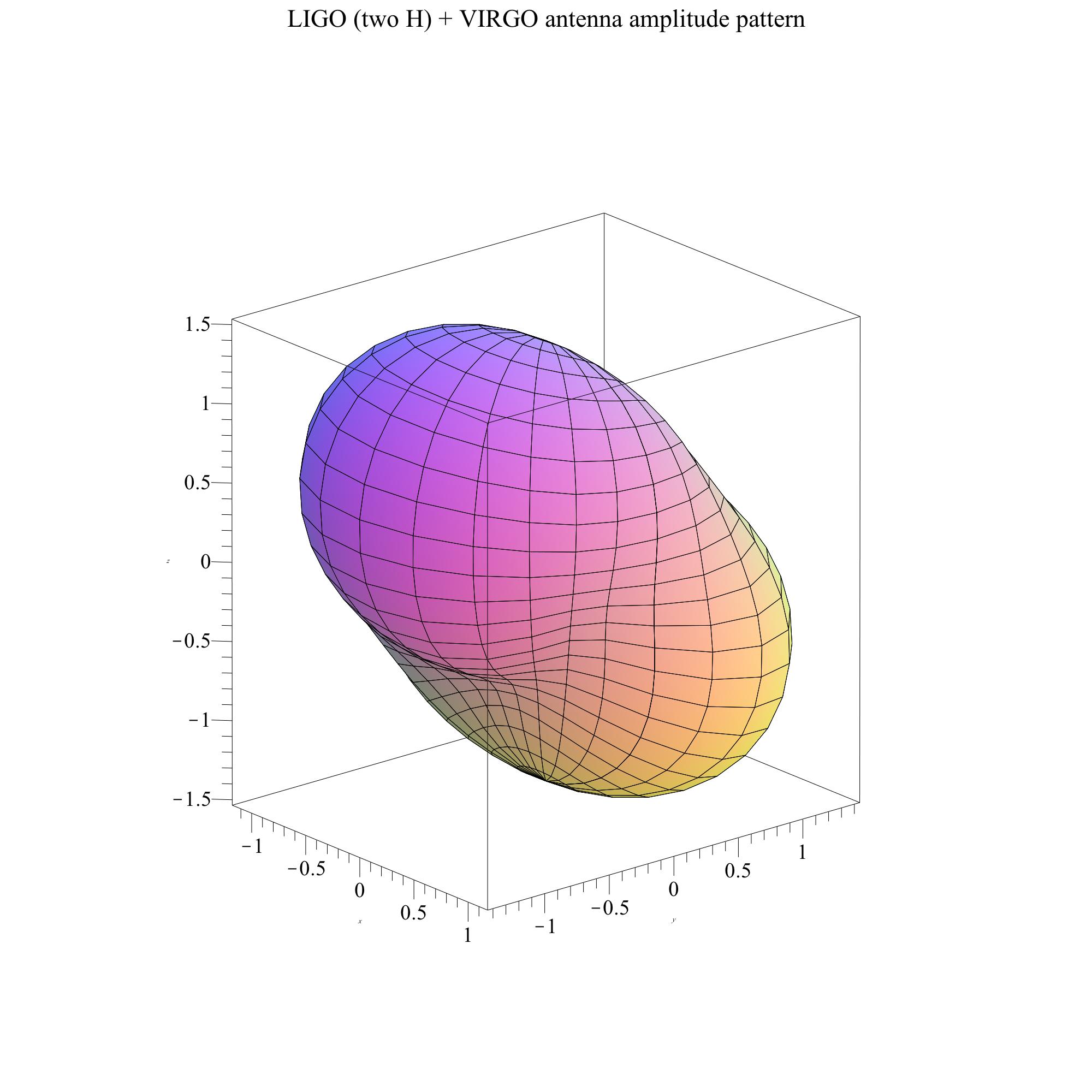}
\includegraphics[width=1.8in]{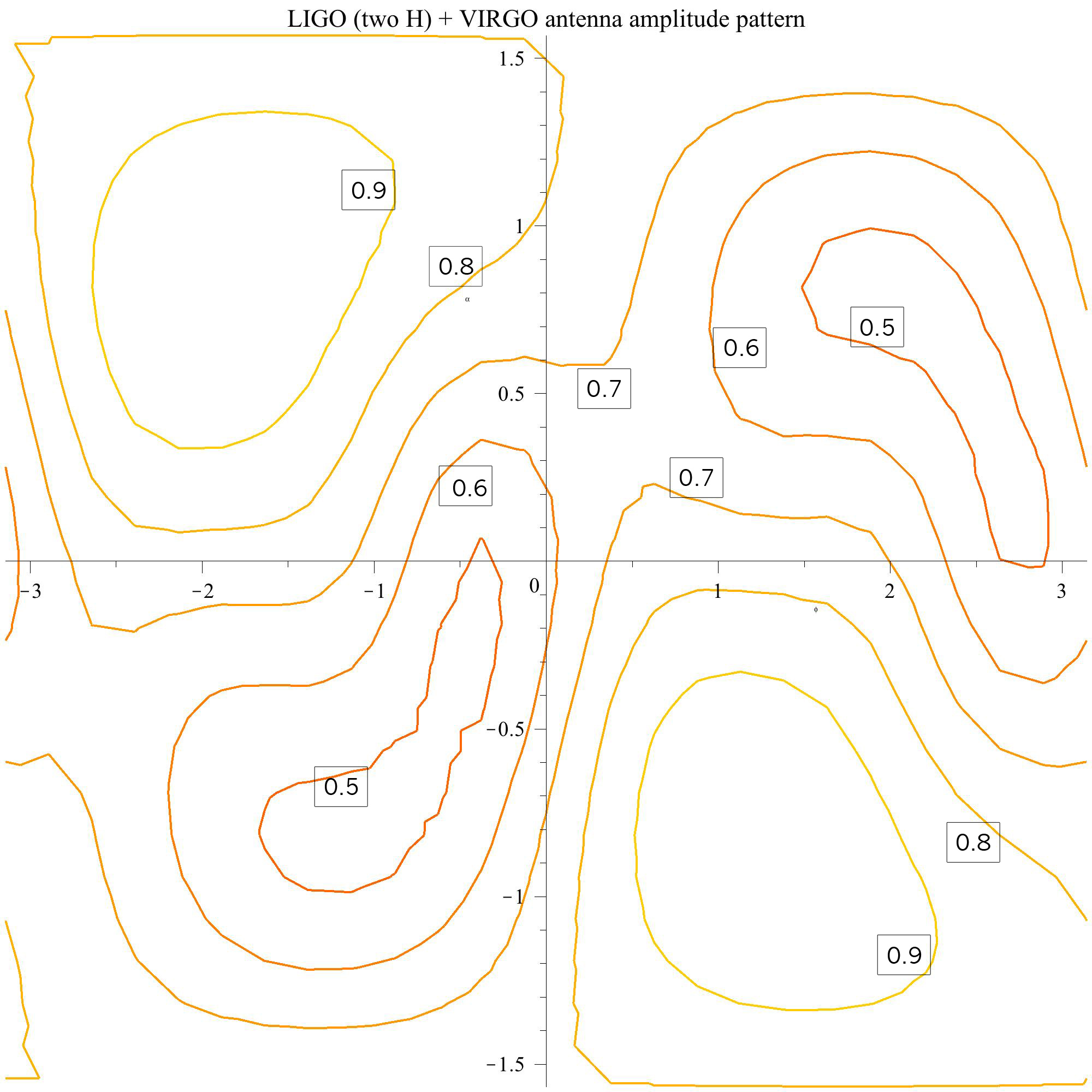}\\
\includegraphics[width=2.4in]{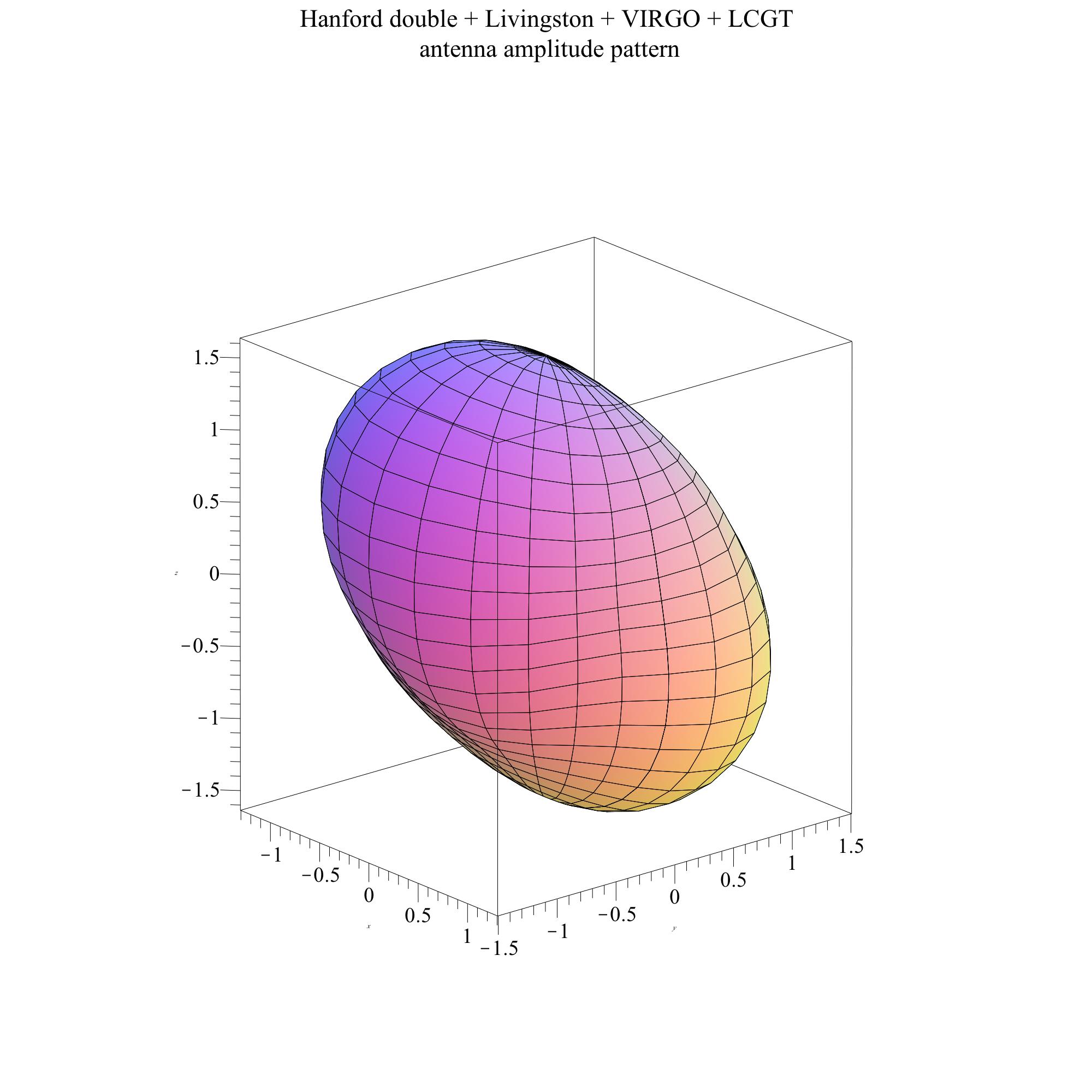}
\includegraphics[width=1.8in]{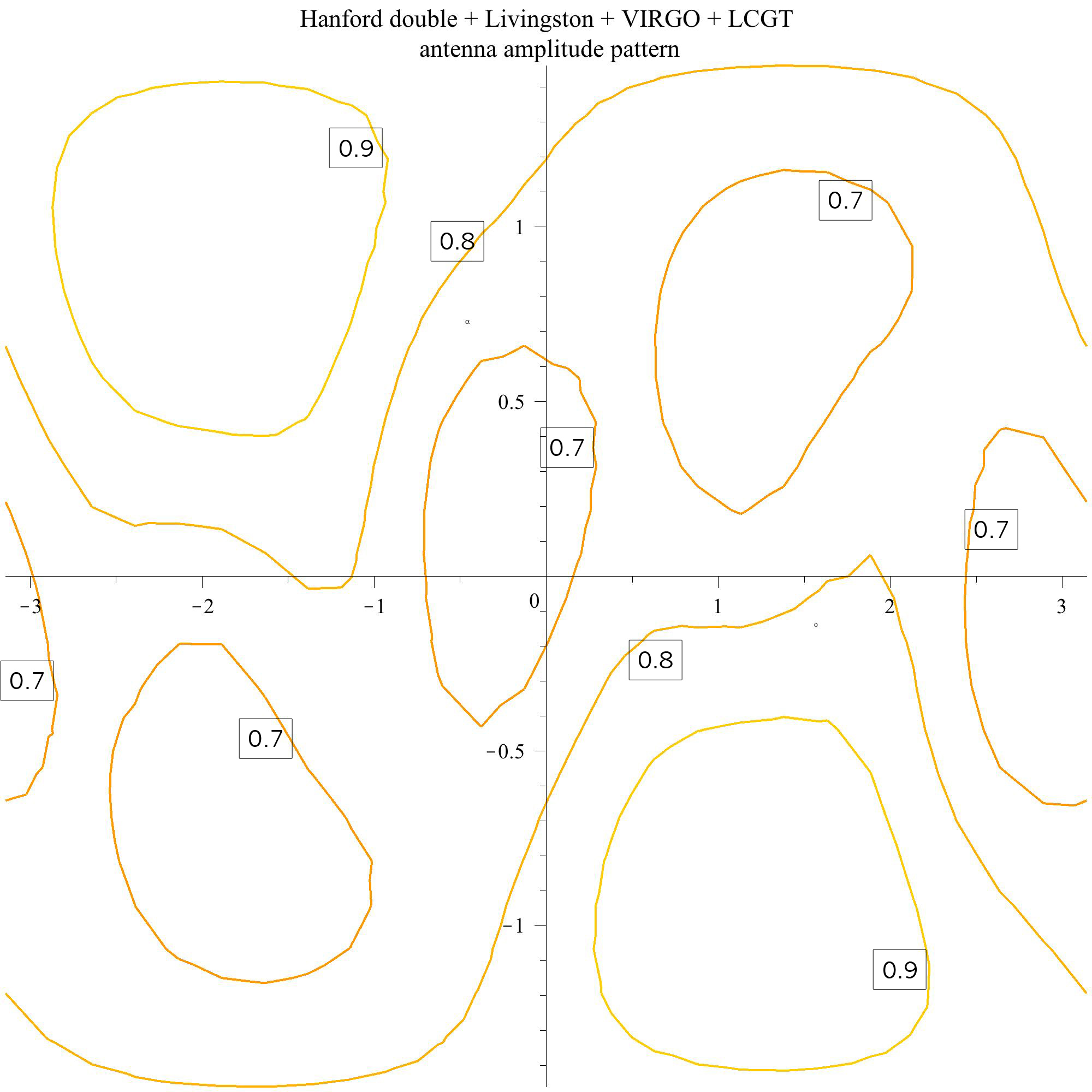}\\
\includegraphics[width=2.4in]{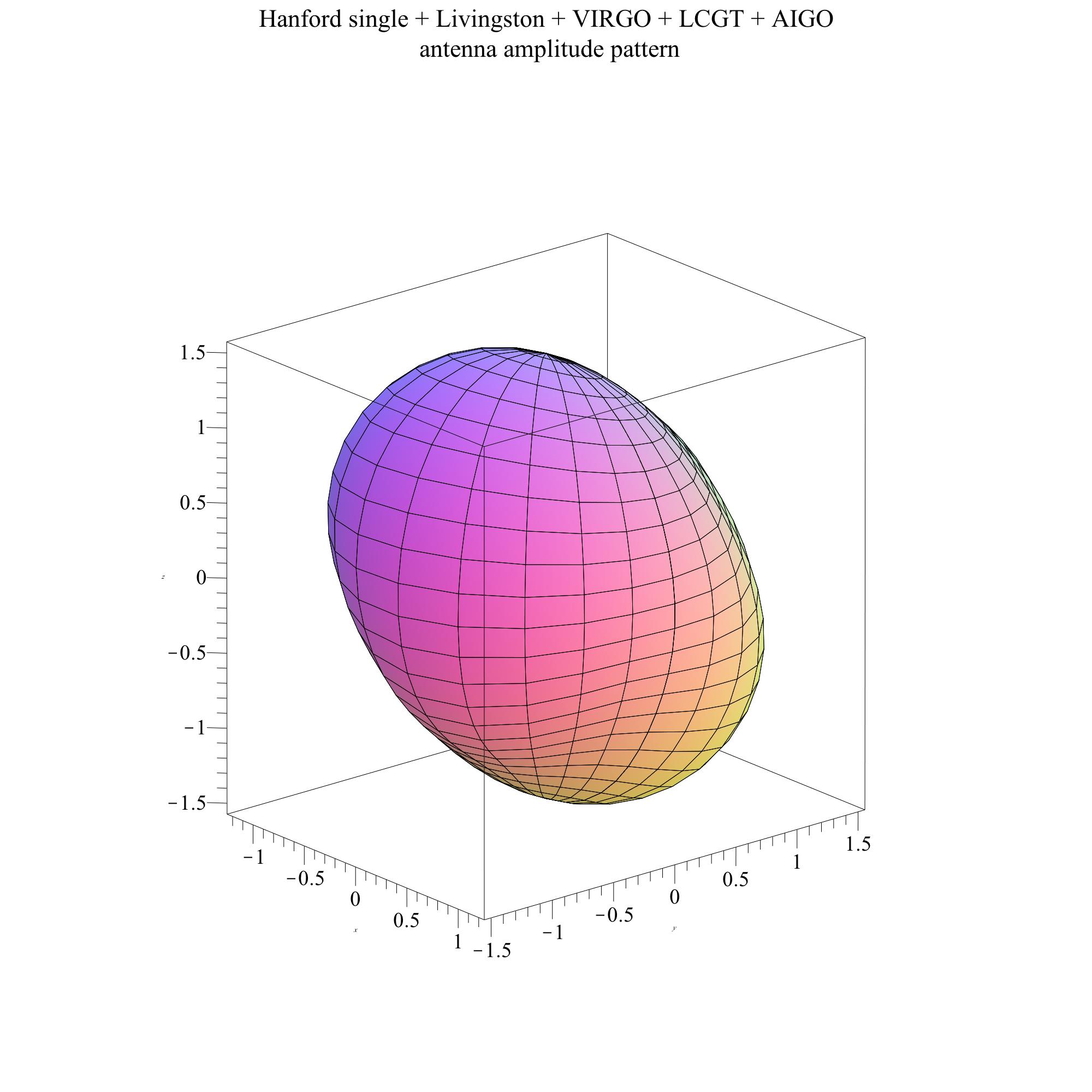}
\includegraphics[width=1.8in]{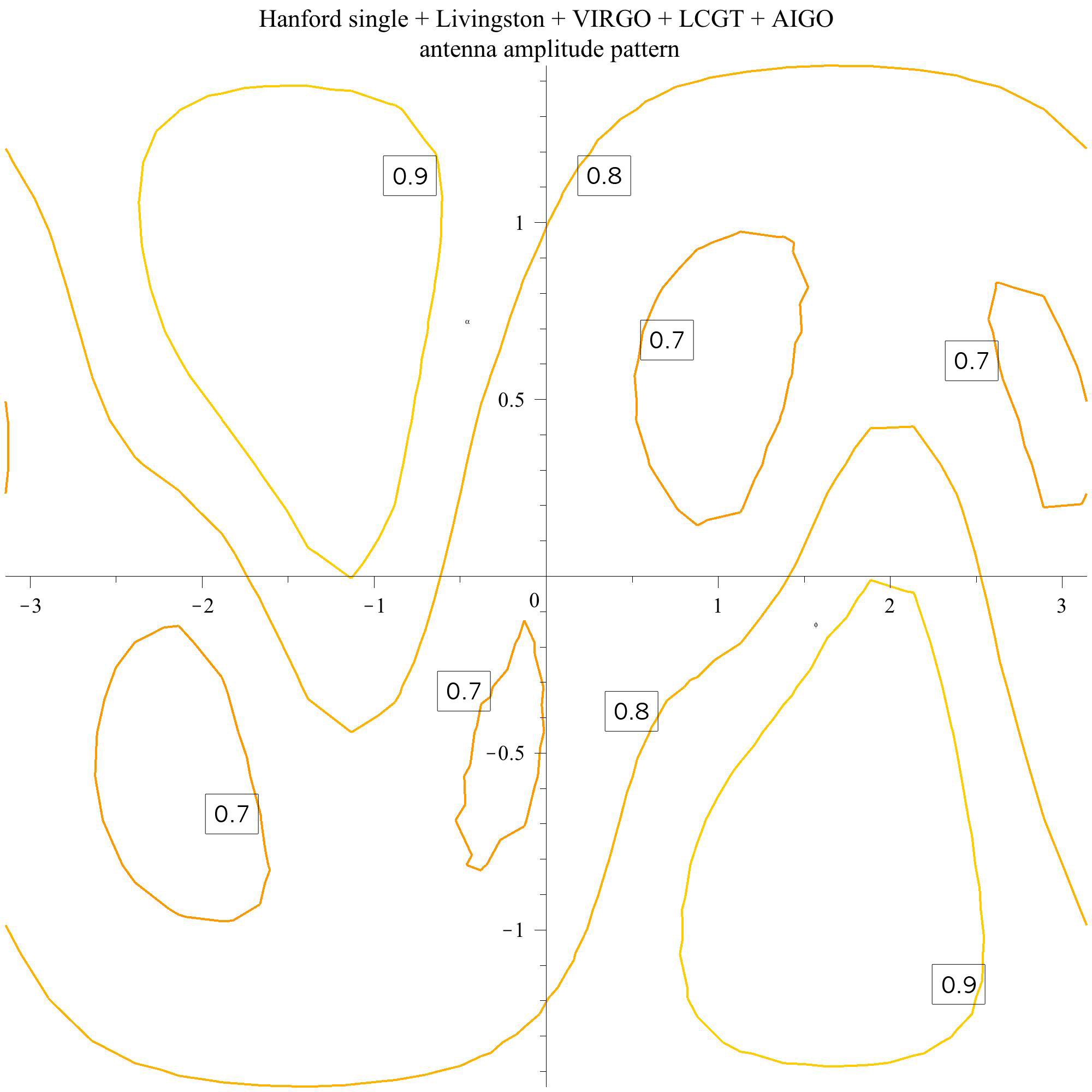}
\caption{Three network amplitude patterns, which show the true spatial shape of the detection volumes. As in \fref{fig:power.HHLV.HHLVJ}, two views are shown, one in perspective and the other as a contour plot. The networks are: (top row) the basic network of two instruments at Hanford, one at Livingston, and one at Pisa; (middle row): the basic network with LCGT in Japan added; (bottom row) the same after moving one of the Hanford detectors to Australia. Notice that all these networks have roughly the same maximum range (HHLV: 1.74; HHJLV: 1.82; AHJLV: 1.76), and these are the values to which the contour levels are scaled. They have different volumes (HHLV: 8.98; HHJLV: 12.1; AHJLV: 12.1) because of their different isotropy, shown in \fref{fig:skycoverage}. (The numbers are taken from \tref{tab:networks}.)}
\label{fig:amplitudepatterns}
\end{center}
\end{figure}

\begin{figure}[htbp]
\begin{center}
\includegraphics[width=3in]{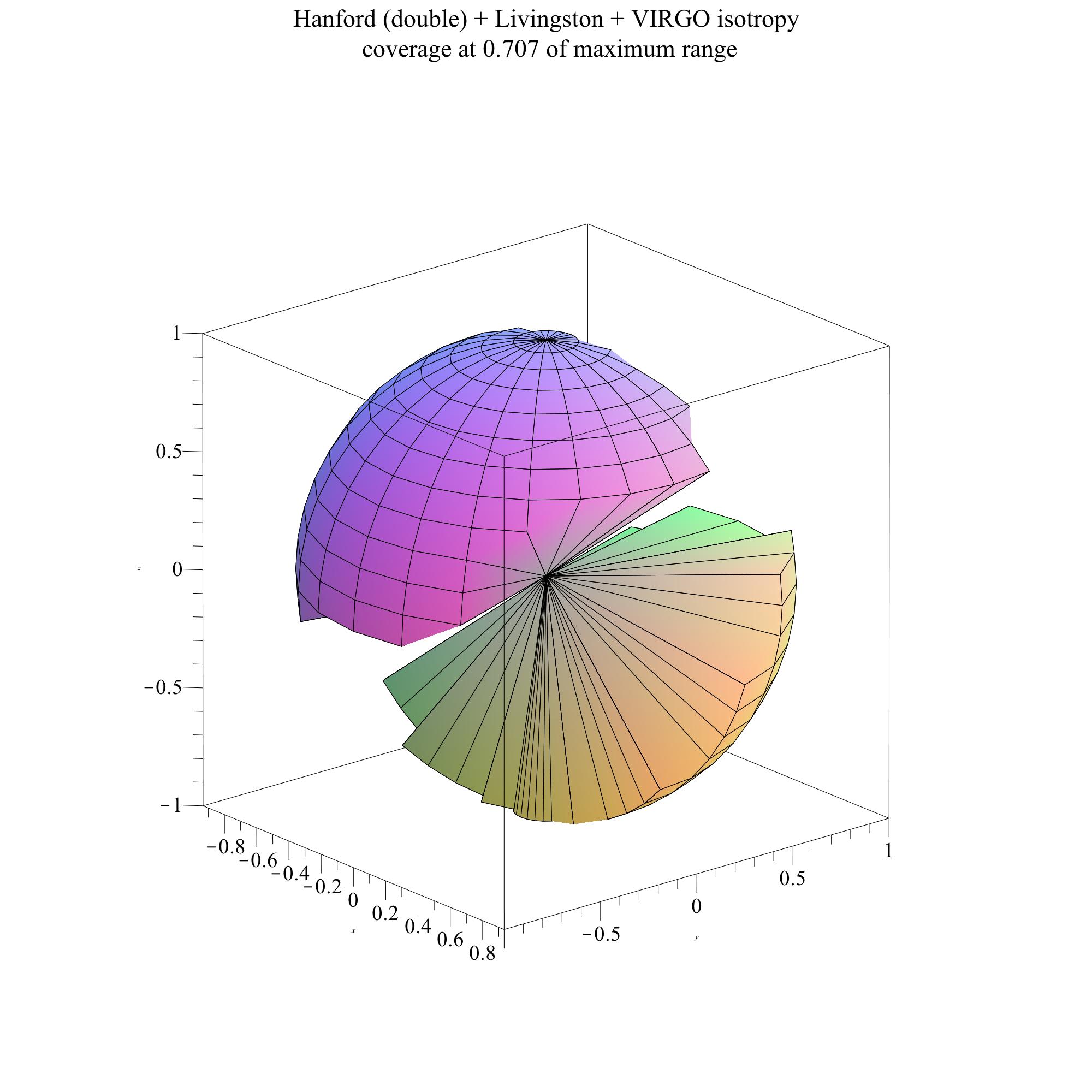}
\includegraphics[width=3in]{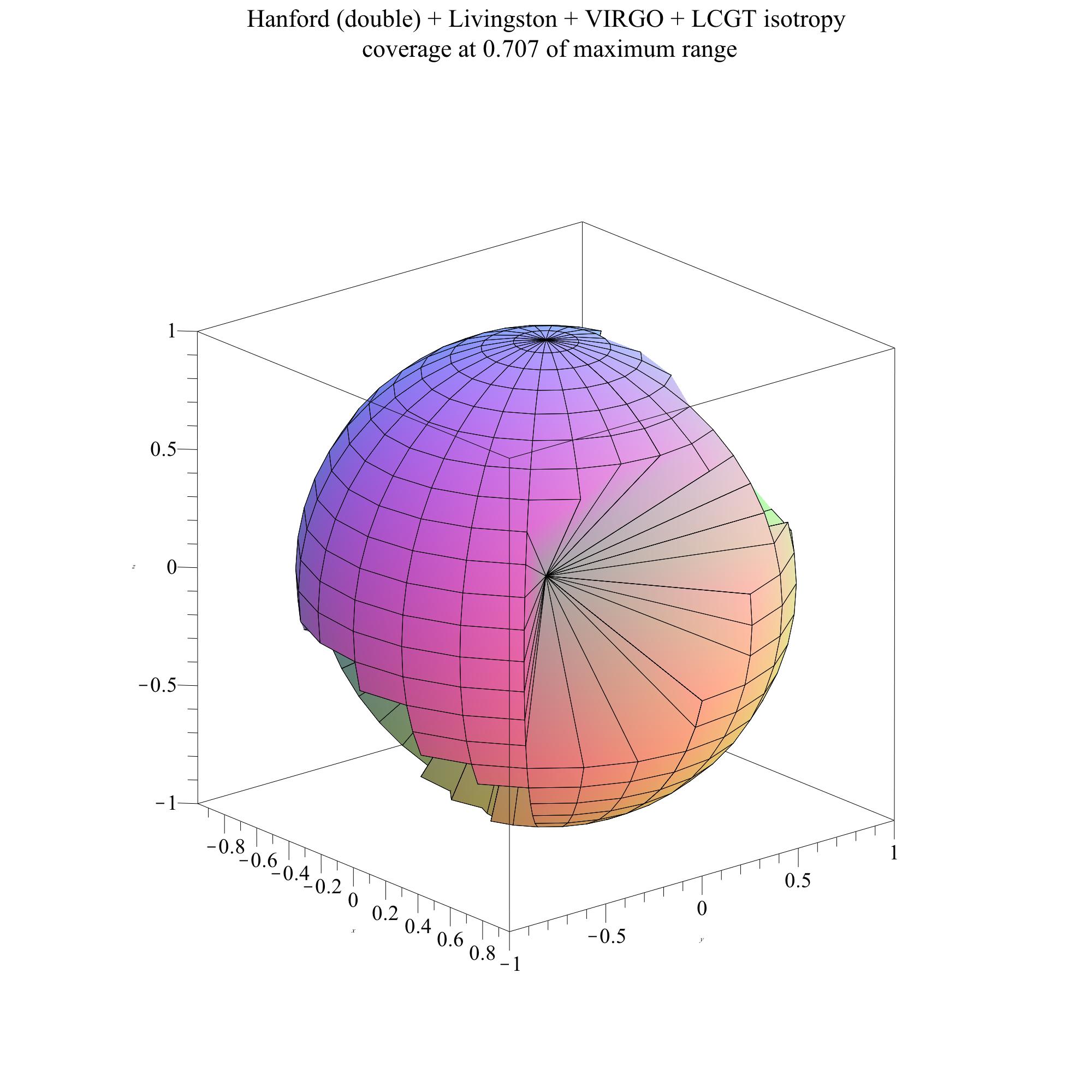}
\includegraphics[width=3in]{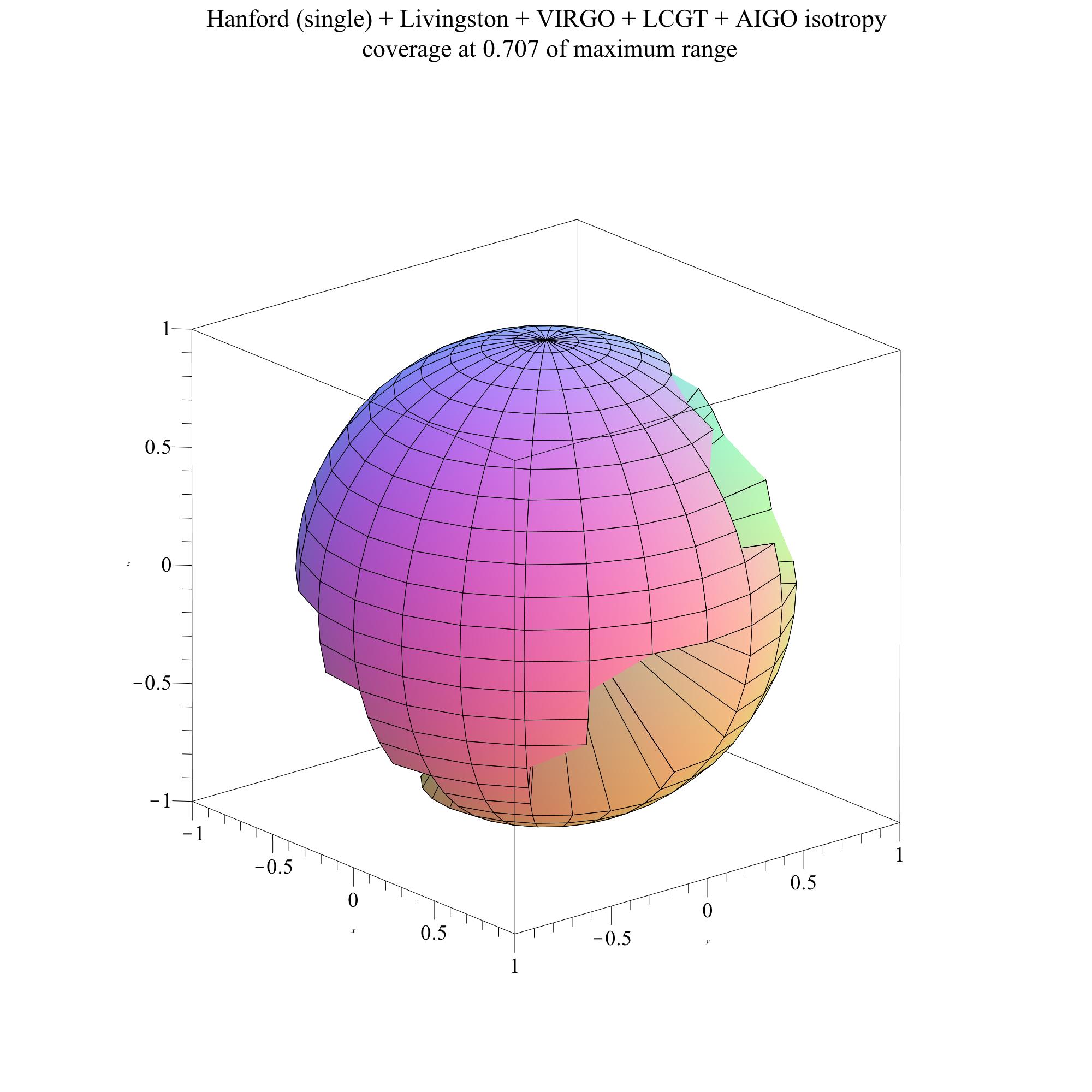}
\caption{Three network isotropy patterns, which show the parts of the unit sphere where the amplitude sensitivity of the detector is better than $\sqrt2$ of its best sensitivity. The networks are the same as in \fref{fig:amplitudepatterns}. }
\label{fig:skycoverage}
\end{center}
\end{figure}

Next we examine the improvements brought by the LCGT detector in Japan, with the simplifying assumption that it will have identical sensitivity to the other Advanced detectors. If there is no detector in Australia then we will have the network HHJLV. Its overall detection volume, at 12.1 (\fref{fig:amplitudepatterns}), is significantly greater that that of HHLV (8.98) and AHLV (8.93), reflecting the fact that there is one further detector. The improvement in the detection rate as measured by \foma\ is even greater: with a Japanese detector and duty cycles of 80\% the rate of detection would be more than 70\% higher than for the basic HHLV, and more than a third higher than AHLV. The network is also significantly more isotropic as well, with \fbshort\ at 73.5\% (\fref{fig:skycoverage}). Adding the baseline to Japan also greatly improves the direction-finding, although not by as much as the longer Australian baselines would: for HHJLV the value of \fcshort\ is 2.57, much better than the 0.66 turned in by HHLV but a bit below the 3.01 value of AHLV. Nevertheless, the improvement over the basic HHLV still represents a 4-fold reduction in the typical area of the error ellipses. 

The Japanese detector may instead operate with a LIGO detector in Australia. To see the difference with the characteristics we found in the previous paragraph, we compare AHJLV with HHJLV.  In detection volume and event rate the two networks are essentially indistinguishable (\fref{fig:amplitudepatterns}). Sky coverage goes up a noticeable amount with the Australian option, from 73.5\% to 85\% (\fref{fig:skycoverage}). And, as might be expected, the extra baselines to Australia and between Japan and Australia improve the direction finding. The value of \fcshort\ for AHJLV is 4.24, compared with 2.57 for HHJLV. So also here the improvement in angular position information provides a strong reason for putting the LIGO detector in Australia. Conversely, if one takes the Australian detector as a given and asks what improvement is brought by the detector in Japan, the comparison is between AHJLV and AHLV. Here not only is direction-finding significantly better (4.24 compared to 3.01), but there is a dramatic increase in isotropy (from 53.5\% to 85\%) and a factor of 1.4 increase in event rate (from 6.06 to 8.71 at 80\% duty cycle). 

On the basis of these numbers the network with the Australian option and the LCGT instrument in Japan looks close to the ideal use of the resources being invested by the various countries involved. It will have nearly twice as many detections per year as the basic HHLV would if it could operate in coherent detection mode (see below), at 80\% duty cycle. It will cover nearly twice the sky area. And its typical direction error ellipses can be a factor of 6 smaller in area. These benefits are brought simply by building one further detector in Japan and moving a detector from the US to Australia.

A nascent project in India might also succeed in building a detector. I have included it in networks by placing it rather arbitrarily at the site of the Giant Metrewave Radio Telescope (GMRT) radio telescope. It is interesting to ask what the properties of networks containing this detector would be. I include the Japanese detector and consider the two LIGO options: HHIJLV and AHIJLV. Adding the Indian detector to the existing HHJLV network increases the event rate by roughly 1/3, regardless of duty cycle. Considering that this is achieved by adding one detector to a network of 5, which is an investment of 20\% on top of the existing expenditure, getting a return of 33\% in terms of science still makes a strong case for this development. The detector in India also improves isotropy, from 74\% to 91\%. And the extra baselines improve position error ellipses, as measured by \fcshort, by 30\%. If the Australian detector is also built, then we compare AHJLV with AHIJLV. Again the Indian detector brings an improvement of around 1/3 in event rate and it achieves nearly complete isotropy, with a value of \fbshort\ of 95\%. It brings a 15\% improvement in position error ellipses, as measured by \fcshort, simply by adding more baselines to the network. 

Several of these networks have recently been studied also by Fairhurst \cite{2010FairhurstLocalization}, who concentrated on the localization ability, using a different approach than that adopted here, and one that is closer to the present methods of data analysis based on thresholding. His results on comparisons of the localization abilities of different networks are broadly in agreement with the relative values of \fcshort\ in \tref{tab:networks}, and the typical ellipse areas that the present treatment gives using \eref{eqn:fomctoaccuracy} are within factors of two of the typical values obtained by Fairhurst. This gives us confidence that our figures of merit can be used not only to compare networks but also, to within factors of two, to characterize the performance of individual networks. 

Another useful comparison is between our analytic results and the Monte-Carlo simulations for coalescing binaries performed by Nissanke, {\em et al} \cite{0004-637X-725-1-496}. They take HLV as their baseline network, i.e.\ assuming only one detector at Hanford, and they do not allow for duty cycle down-time. They find that AHLV will detect 1.48 times more events than HLV. In \tref{tab:networks} the appropriate comparison is between the full detection volumes of HLV and AHLV, whose ratio is 1.55. We take this to be excellent agreement. Moreover, they measure the isotropy of various networks by plotting detected event distributions on the sky (their figure~2). Their conclusions are qualitatively in agreement with ours in \fref{fig:skycoverage}, and they remark that networks that include LCGT are noticeably more isotropic, a conclusion also in agreement with our values of \fomb\ in \tref{tab:networks}.

Note, however, that the true ``default'' network is HHLV, and in  \tref{tab:networks} it is clear that moving one of the H detectors to Australia hardly changes the total detection volume. When network duty cycle is taken into account, there is a net event rate gain (for three-site detections) of up to a factor of $1.24$. On top of that there is an event rate gain from being able to do coherent data analysis better, so that the LIGO Australia option not only has better angular resolution but also a significantly higher detected event rate. This is the subject of the next section.

\subsection{Coherent versus coincidence data analysis: implications for event detection rates}\label{sec:cc}

The assumption of this paper is that data analysis is done by fully coherent combination of the different detectors' data streams. This is not yet the practice in the LSC-VIRGO data analysis, mainly because coherent analysis normally assumes a Gaussian background of instrumental noise, and is therefore vulnerable to what are often called ``glitches'', bursts of noise from instrumental effects that can masquerade as real signals. Because in present detectors there is a significant glitch background, data analysis usually includes a {\em coincidence} step, in which events of a sufficient size in single data streams that occur in coincidence (within a time-window equal to the light travel times among the various detectors) with events in other detectors are selected and studied further. This coincidence test eliminates most of the glitch background. 

But a purely coincident analysis also eliminates most of the potentially detectable signals, i.e.\ signals that could reliably be detected if the background noise were ideally Gaussian. The penalty is easy to compute. In a recent review of the astrophysical evidence for the rates of compact object binary coalescences, the LSC and VIRGO collaborations predicted a detected event rate for the HHLV network of Advanced detectors \cite{2010CQGra..27q3001A_short}. Their method was to take the number of events that occur inside the detection volume of a {\em single} detector above the detection threshold $\rho_{\rm min}=8$. They took the most likely value of the rate of neutron-star coalescences to be $1\,\rm {Mpc}^{-3} \, {Myr}^{-1}$, or equivalently 100 events per Milky Way Equivalent Galaxy per million years. With this volume event rate, the most likely detection rate for these systems came out to be 40 per year. The reason for counting only events that occur in one detector's detection volume despite the fact that the network contains four detectors is to approximate in a rough (and conservative) way the coincidence criterion. 

\begin{figure}[htbp]
\begin{center}
\includegraphics[width=6in]{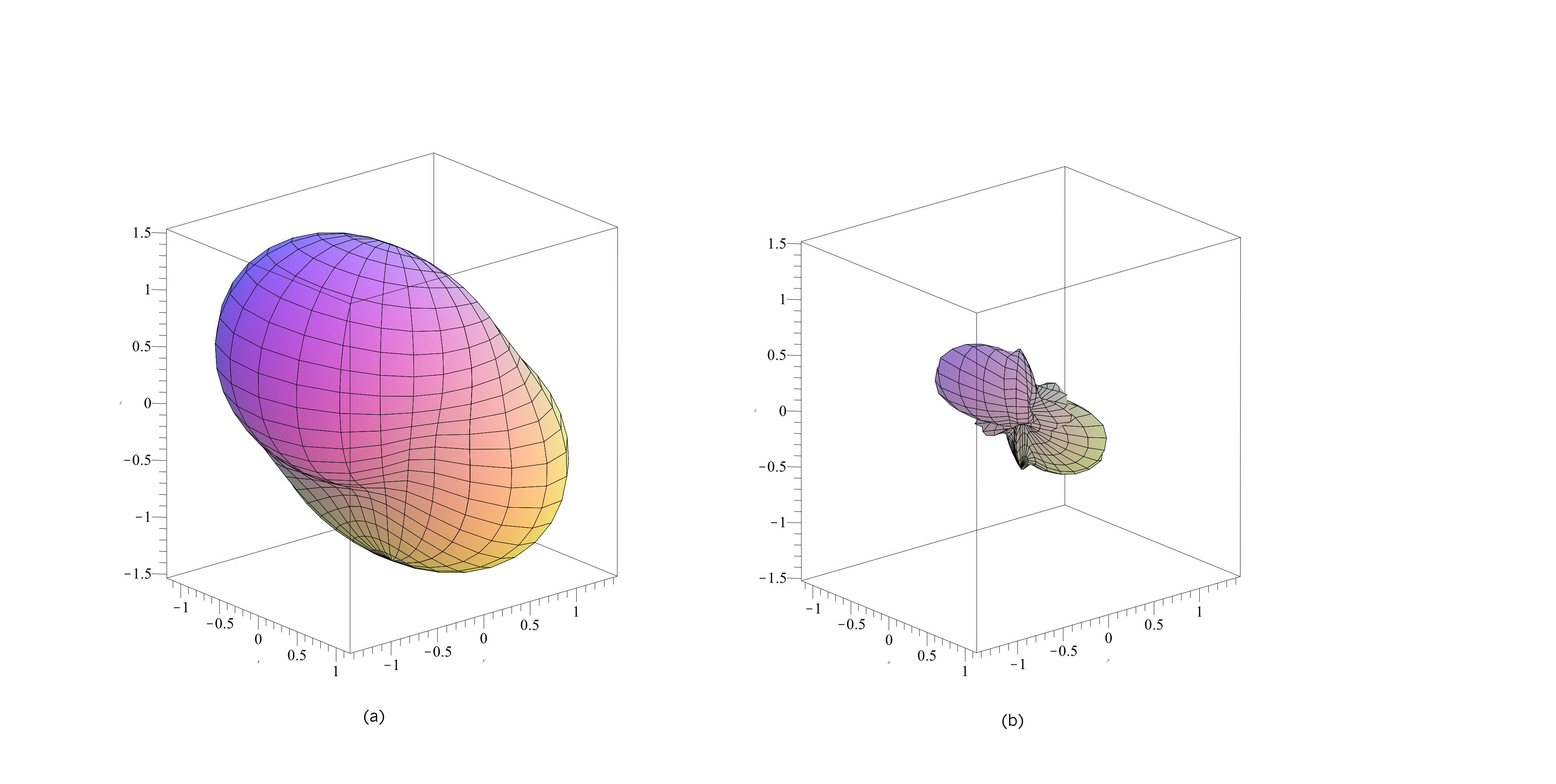}
\caption{The antenna patterns of the LIGO-VIRGO detectors for (a) coherent and (b) coincidence analysis methods. The coherent pattern is the HHLV amplitude pattern. The coincidence pattern is the region in which, for random polarizations, an event crosses threshold in at least two of the detectors (but not allowing events that appear only in two Hanford detectors). The thresholds are assumed to be the same, e.g.\ if the individual detector thresholds for the coincidence analysis is 8, then the coherent data analysis threshold is also set at 8, as discussed in the text.}
\label{fig:compare}
\end{center}
\end{figure}

For the same network, but with coherent data analysis using a network threshold of the same value ($\rho_{\rm N,\,min}=8$.), the data in \tref{tab:networks} show that the rate would be higher by the ratio of \fashort\, which is 4.86 (allowing for an 80\% duty cycle), to the volume for a single detector, 1.23. This ratio is 3.95, which implies that the HHLV network, with perfectly Gaussian noise, could detect about 160 events per year if it did coherent analysis. {\em The difference in detection effectiveness between coherent and coincidence analysis for coalescing binary signals in this basic network is a factor of 4 in detection rate. } This difference is illustrated graphically by comparing the volumes of space covered by fully coherent analysis and pure coincidence analysis, in \fref{fig:compare}.

Naturally this comparison depends on the threshold assumed for the two kinds of data analysis. The comparison shown in the figure is for equal thresholds: if, as in  \cite{2010CQGra..27q3001A_short}, the coincidence observation is done with a threshold SNR of 8 in each detector, then we assume that the network coherent threshold is set at 8 as well. This is not unreasonable, since the coherent analysis essentially fights only against Gaussian noise, where events at $8\sigma$ occur only once in $10^5$ years at an effective sampling rate of 300~Hz. This works well if coherent methods can eliminate glitches. This may not be fully possible for HHLV (see below) but it should be possible for the enlarged networks, including AHLV. Therefore the comparison shown in this figure is relevant for extrapolations of event rates to the larger networks.

Now, for the existing detectors, instrumentalists are working hard to reduce the glitch rate, and the LSC-VIRGO analysis teams are bringing in coherent analysis \cite{2007PhRvD..75f2004R,Klimenko:2008fu,2008CQGra..25r4025M,2008CQGra..25r4016H,2010PhRvD..81f2003V}. Networks containing three or more detectors can also use their null streams to test for and veto glitches, as described in \sref{sec:netcohere}.

In practice, the analysis teams will begin by mixing coincidence and coherence methods, by setting a low threshold on coincidences to obtain a population of possible events, and then using coherent methods (including null streams) to eliminate the glitch coincidences. Such methods are computationally much less demanding than purely coherent methods, and they can presumably bridge some of the gap of the factor of four between pure coincidence and full coherent methods. 

However, the basic HHLV network may not be completely amenable to coherent analysis, because of the near-perfect alignment of the LIGO Hanford and LIGO Livingston detectors. While this allows good discrimination against glitches in one of the LIGO detectors, it reduces the information recoverable from real events: polarization and sky location can be determined only if VIRGO is excited comparably strongly to the LIGO detectors, and without a sky location one cannot define a null stream. This in turn leads to more opportunities for false alarms, and lowers the significance of real events. It remains to be seen how much of the full factor of 4 computed above can be recovered by introducing some degree of coherent analysis into the HHLV network, but clearly it is a very important step to take. If the nominal detection rate of 40 events per year can be raised even to 80, this will be the most cost-effective way to improve the baseline network. 

It is worth noting that the LSC study of the LIGO Australia option \cite{LIGOsouth} made a strong recommendation to move to coherent data analysis. The move of one LIGO detector to Australia breaks the degeneracy of the LIGO instruments, especially if the new detector is anti-aligned with the existing LIGO detectors. It should therefore allow fully robust coherent analysis, coming close to the maximum possible event detection rate of 200 NS-NS events per year, assuming the most likely rate quoted in \cite{2010CQGra..27q3001A_short}, and assuming an 80\% duty cycle. {\em The improvement of a factor of up to 5 in the detection rate is probably the strongest reason for placing a LIGO detector in Australia.}

The LCGT detector will add a third null stream to the HHLV or AHLV networks, and make coherent analysis even more robust. If the most likely coalescence rates prove to be accurate, and if the network detection threshold is set to 8, the HHJLV network can expect to detect 270 NS-NS coalescences per year, and the AHJLV network 280. Adding a detector in India raises these numbers to around 370 events per year. Improving the duty cycle to 95\%, which seems feasible after a few years of operation, increases the five-detector NS-NS rates to around 360 per year and the six-detector rate nearly to 500 per year. {\em These rate improvements would qualitatively change the kind of science obtainable from Advanced detectors.}

For coalescences of neutron stars with black holes, the LSC and VIRGO paper \cite{2010CQGra..27q3001A_short} quotes a ``best'' rate of 10 per year for HHLV with coincidence analysis. The expected rates for larger networks can therefore be obtained from the NS-NS rates just quoted by dividing by 4. Similarly, the rates for binary black hole mergers are expected to be half of the NS-NS rates; black holes have a much lower number density in the universe, but they can be detected much further away because of their higher mass. The NS-BH and BH-BH rates are, of course, much less secure than the NS-NS rates, because there are no observed binary systems of those types; the rates used in  \cite{2010CQGra..27q3001A_short} depend exclusively on population simulations. However, the recent identification of two possible X-ray binary precursors of BH-BH binaries provides a much-needed observational normalization of the population. Bulik {\em et al}  \cite{2011ApJ...730..140B} conclude from these systems, in which a black hole is in a close binary with a Wolf-Rayet star, that the BH-BH detection rate might in fact be much higher and could even significantly exceed the NS-NS rate.

One further item is worth noting. Searches for binary signals are optimal if they incorporate as much prior information as possible, and Bayesian analysis techniques that do this are becoming standard in the current LSC-VIRGO data analysis methods. The present study provides three such priors: the network antenna pattern (a prior on the sky location of the source) and the two p.d.f.'s: the expected distribution of SNR values \eref{eqn:snrpdf}, which is a prior on the signal amplitude; and (for binaries) the expected distribution of detected inclination angles (\eref{eqn:pdfiota} and \fref{fig:iotabias}), which is a prior that affects the relative amplitudes and phases of the signal in different detectors. The use of the antenna pattern as a prior needs to be done with care, because as noted above there will be a number of sources detected that are outside the ``hard'' edge of the detection volume. A polarization-dependent prior is of course even better than the polarization-averaged antenna pattern computed here.

\section{Conclusions}\label{sec:conclusions}

In this paper I have developed a framework in which it is possible to compare networks of gravitational wave interferometers consisting of different numbers of detectors in different geographical configurations. I have shown that, for any network, the expected SNR distribution of detected events, once the data analysis can be done by optimal coherent methods, is a universal $\rho^{-4}$ power law that falls to zero for $\rho$ smaller than the detection threshold. It follows from this distribution that the most likely SNR of the first detected signal will be about 1.26 times the threshold of the search. I have derived the (similarly universal) probability distribution of the inclination angle of detected binary systems, and I have shown that, if coalescing binaries are associated with narrowly beamed gamma-ray bursts, then because the radiated gravitational wave power is correlated with the direction of the gamma-ray cone, we can expect 3.4 times more detected coincidences than if they were not correlated.  I have suggested three figures of merit that can be computed for any network and which measure average properties of the network: its expected event detection rate, its isotropy, and the accuracy of its sky position measurements. These figures of merit are inevitably crude averages, and they should not be a substitute for detailed comparisons of networks as part of the planning for specific new detectors. But they give a clear indication of the merit of enlarging the network from the originally planned LIGO and VIRGO detectors to include detectors in Asia and Australia. 

It is worth stepping back from the many different options that exist for enlarging the worldwide interferometer network to consider the net improvements that are possible if current plans are realized. Consider the network AHJLV, consisting of LIGO with one detector in Hanford and one in Livingston, VIRGO in Italy, LCGT in Japan, and LIGO Australia. The numbers in \tref{tab:networks} show how much more science that network can do than the originally planned HHLV. Its event rate, with detectors operating on 80\% duty cycles, would be nearly twice as high for all categories of burst sources. It would cover nearly twice as much of the sky, making it a better bet for coincidence observations with neutrino detectors. And our measure of the areas of angular position measurement error ellipses improves by a factor of 6.4, from 0.66 to 4.24, indicating that the typical error ellipse goes down in area by a factor of more than 6. This will make a huge improvement in follow-up studies with optical and other telescopes. This network offers much more science than had been promised in the initial proposals for the existing four large detectors, at the cost of building only one more detector and moving another to a better location. The impact of the single extra detector in Japan is so large because robust gravitational wave astronomy requires a minimum of three detectors in different locations, so the marginal impact of increases to four and five is large.

If the project in India gains support and, on a longer timescale, leads to a sixth Advanced detector, it would create the network AHIJLV, an even bigger improvement on HHLV. Its event rate would be 2.4 times higher, on a duty cycle of 80\%. It would cover 95\% of the sky at half power, and its sky localization error ellipses would be fbetter than 7 times smaller in area than those of the presently planned LIGO-VIRGO network.

It is important to realize that both of these enlarged networks have maximum detection distances that are within 5\% of the maximum range of HHLV. Their large event rate gains come partly from increased isotropy and partly from having more three-site sub-networks that can detect and localize events even when one or more detectors has fallen out of observing mode. They survey the same volume of space more completely than HHLV can. But the big improvements in sky localization are perhaps the strongest arguments for pursuing these enlarged networks. The values of \fomc\ we compute here suggest (using the conversion to steradians given above) that the typical error box in either network would be smaller than a degree on a side. This not only makes searching with electromagnetic telescopes for counterparts easier but it reduces the probability of chance coincidences in a large field of view.

The conclusions in this paper depend strongly on the assumption of coherent data analysis. If coincidence data analysis is used, where events are selected for further study only if they cross a particular threshold in each participating detector, there is no guarantee that the properties described here will still hold for the different networks. Coherent analysis produces networks whose antenna patterns are the sum of the power patterns of the network members. Coincidence analysis produces antenna patterns that are basically determined by the intersections of the power patterns of network members. Performing a first cut at the noise by coincidence analysis, even if it is followed by a coherent follow-up, will not reproduce the assumptions used here. The reason for coincidence analysis is, of course, to eliminate rare but strong non-Gaussian noise events, but these can also be identified by using network null streams, whose number increases with the number of detectors in the network.

Moving from coincidence to coherent analysis can increase detection rates by factors of four or more. It is to be expected that network data analysis will move to fully coherent analysis as the number of detectors increases and as experimenters manage over time to reduce the frequency and amplitude of non-Gaussian noise glitches. With such analysis techniques,  the full potential of the enlarged networks, as illustrated by the figures of merit calculated here, can eventually be realized.

\ack 
It is a pleasure to acknowledge discussions of the network problem with many colleagues, including B Krishnan, S Dhurandhar, J Hough, K Kuroda, A Lazzarini, and J Marx. Special thanks to S Klimenko, S Nissanke, M-A Papa, P Sutton, B Sathyaprakash, and L Wen for detailed discussions and comments. This work was stimulated by a kind invitation from K Kuroda to the 58th Fujihara Seminar in 2009. DFG grant SFB/TR-7 is gratefully acknowledged.

\bibliographystyle{unsrtnat}
\bibliography{networks}

\end{document}